\documentclass[prd,preprintnumbers,superscriptaddress,amsmath,amssymb,nofootinbib,twocolumn,showpacs]{revtex4-2}

\usepackage{dcolumn}% Align table columns on decimal point
\usepackage{bm}% bold math
\usepackage{graphicx}
\usepackage{float}
\usepackage{footmisc}

\usepackage{color}
\usepackage{natbib}
\usepackage{twoopt}
\usepackage{lipsum}

\usepackage[pdftex,
            breaklinks=true,%
            colorlinks=true,%
            linkcolor=red,
            urlcolor=blue,
            citecolor=blue,
            pdfauthor={Par\'e, Lavalle},%
            pdftitle={Template for article in XXX}%
           ]{hyperref}

\graphicspath{{./}{Figs/}}

\usepackage{shortcutLatex}

\begin{document}

\title{Probing the fate of large primordial perturbations with exoplanets}

\author{Th\'eo Par\'e}
\email{theo.pare@umontpellier.fr}
\author{Julien Lavalle}
\email{lavalle@in2p3.fr}
\affiliation{Laboratoire Univers \& Particules de Montpellier (LUPM),
  CNRS \& Universit\'e de Montpellier (UMR-5299),
  Place Eug\`ene Bataillon,
  F-34095 Montpellier Cedex 05 --- France}

% arXiv limit of 1920 characters

\begin{abstract}
We propose ultra-wide-orbit exoplanets as a novel probe of small-scale dark matter objects. These systems are highly sensitive to gravitational perturbations that could be induced by a Galactic population of compact baryon-free dark matter objects---whether point-like or extended. Focusing on ultra-compact minihalos, which may arise from large primordial perturbations deviating from the canonical scale-invariant power spectrum, we derive new constraints on their injection scale and amplitude. These constraints complement existing dynamical limits and are expected to improve with upcoming exoplanet surveys. Furthermore, the detection of additional loosely bound exoplanets with these surveys could significantly tighten these constraints. Beyond constraints, we also identify characteristic observational signatures in these systems that could help trace a population of dark matter objects. All this strengthens the potential of exoplanetary science to probe the dark universe back to its very primordial properties.
\end{abstract}

% 12.60.-i: Models beyond the standard model 
% 95.35.+d: dark matter (stellar, interstellar, galactic and cosmological)
% 95.30.Cq: Elementary particle processes
% 98.35.Gi: Galactic halos
% 97.82.-j: Extrasolar planets
% 98.54.Kt: Primordial galaxies
%\keywords{Suggested keywords}% Use showkeys class option if keyword
                              % display desired
\maketitle
\preprint{LUPM:26-002}
\section{Introduction}
\label{sec:Intro}
The cold dark matter (CDM) paradigm is a compelling theoretical construction built on a minimal set of ingredients and assumptions about the nature of dark matter (DM) and the statistical properties of primordial density fluctuations. It successfully explains, among other key features, the origin and properties of galaxies (see \eg~\cite{Peebles1982,BlumenthalEtAl1984}). This paradigm, where primordial fluctuations are typically assumed to be Gaussian and described by a primordial power spectrum, is supported by a large variety of observations across a wide range of astrophysical scales, most notably the observed statistical properties of the cosmic microwave background (CMB). Similar properties are also observed in the angular distribution of galaxies at later time, as these structures originate from the seeds seen earlier in the CMB. All this is grounded in well-established linear cosmological perturbation theory, valid in the pre- and post-recombination universe before structures formation. In this framework, an exotic collisionless CDM is added on top of standard matter and radiation \cite{MaEtAl1995}. This overall setup can be extended by cosmological simulations in the non-linear regime to study galaxy formation in greater details.\footnote{The linear theory can actually further be used to predict the mass function of CDM structures as a function of time, including their clustering and accretion into bigger structures \cite{PressEtAl1974,BardeenEtAl1986,BondEtAl1991a,LaceyEtAl1993,MoEtAl2010}, a framework that we are going to use later on.} Despite theoretical and practical challenges in modeling baryonic physics on subgalactic scales, the predictions of this framework are in astonishing agreement with observations \cite{VogelsbergerEtAl2019}. Moreover, recent cosmic tensions \cite{AbdallaEtAl2022,Peebles2022,Peebles2025} do not undermine the DM and structure formation aspects of this standard cosmological model.

Observations on cosmological scales are consistent with a nearly scale-invariant spectrum of Gaussian density perturbations, as predicted by the simplest inflationary scenarios \cite{Baumann2009,AchucarroEtAl2022}. This implies that, in the hierarchical picture of structure formation, CDM structures can form down to arbitrarily small scales, with minihalos forming first and later merging into larger systems to become subhalos---if they survive in the process. While a concrete scenario of CDM \cite{CirelliEtAl2024} would introduce a cutoff mass due to kinetic pressure (\eg~free streaming), many particle physics models predict such low cutoff values that CDM structures with masses well below the solar mass are a natural theoretical expectation \cite{HofmannEtAl2001,GreenEtAl2005,BringmannEtAl2007a,Marsh2016}. Notable exceptions include warm \cite{DodelsonEtAl1994,ColombiEtAl1996} and ultra-light DM though \cite{HuEtAl2000,Ferreira2020}. Besides, structures with cosmological masses below $\sim 10^{8}\Msun$ (not tidal masses) are typically too light to efficiently attract baryons and form stars due to UV pressure \cite{VogelsbergerEtAl2019,ZavalaEtAl2019a}. This implies the existence of large populations of completely dark subhalos within larger structures, such as galaxies, in the CDM scenario \cite{SilkEtAl1993,DiemandEtAl2008b,SpringelEtAl2008}. Detecting these objects--directly or indirectly--would be a threefold major discovery: (i) it would definitely prove CDM, (ii) it would tightly constrain the nature of dark matter (DM), and (iii) it would probe the properties of primordial perturbations on unprecedented small scales.

In fact, it is not unreasonable to expect departures from scale invariance on small scales within or beyond single-field inflation models \cite{Starobinskii1992,BaumannEtAl2014,BallesterosEtAl2018a,ByrnesEtAl2019,AchucarroEtAl2022}, or due to other mechanisms such as phase transitions \cite{FrancioliniEtAl2026}. Extreme amplification of primordial perturbations is the primary production mechanism for primordial black holes (PBHs), which would themselves act as a CDM component and respond gravitationally to fluctuations on scales larger than their own seeds \cite{CarrEtAl1974,Chapline1975,CarrEtAl2020b,CarrEtAl2021d}. More moderate yet sizable amplification of small-scale perturbations relative to the canonical scale-invariant lore can also produce very dense CDM structure known as ultra-compact minihalos (UCMHs) \cite{BerezinskyEtAl2003,RicottiEtAl2009,BerezinskyEtAl2010,BerezinskyEtAl2011,BerezinskyEtAl2014,DelosEtAl2018a,DelosEtAl2023}.

Depending on the nature of DM, dark minihalos could give rise to various signatures. From an agnostic perspective though, the most generic ones arise from their gravitational interactions with visible matter. Among these, the dynamical heating of visible systems has long been used to probe dark matter objects (DMOs) in the Milky Way and other galaxies (\eg~\cite{LaceyEtAl1985,JohnstonEtAl2002,PenarrubiaEtAl2016,Penarrubia2018,ChiangEtAl2023,GrahamEtAl2026}). Complementary gravitational probes include, for instance, lensing \cite{GilmanEtAl2026,CroonEtAl2026,BringmannEtAl2025,Ando2026} and pulsar timing \cite{DrorEtAl2019}.

In this work, we propose exoplanetary systems as a new dynamical probe of CDM substructure, modeled as spherically symmetric DMOs orbiting the Milky Way. The compactness and mass of these DMOs depend on the underlying theoretical assumptions about the amplitude of primordial perturbations. While exoplanets have been used in DM searches---primarily to study their ability to capture dark matter particles and related phenomena \cite{BenitoEtAl2024,PhoroutanMehrEtAl2025,IlieEtAl2024}--dynamical aspects have been less explored. A few exceptions arise in studies of PBHs---see \eg~\cite{LehmannEtAl2022,HalpernEtAl2025}---but not related to the present study. Here, we essentially exploit exoplanetary systems in a manner similar to wide stellar binaries have been used to probe dark compact objects \cite{PenarrubiaEtAl2010a,PenarrubiaEtAl2016,RamirezEtAl2023a,TylerEtAl2023,ShariatEtAl2025,BhallaEtAl2025}. However, wide stellar binaries may suffer from uncertainties in their age determination, and, for the sake of speculation, from the possibility that they could even form from encounters with a DMOs in the first place. Exoplanetary systems could be less affected by these issues, as the age of a planet and that of its parent star are nearly identical \cite{Crida2023,Mordasini2018,DeegEtAl2026}. This admittedly makes them cleaner proxies for ``binary'' system to probe DMOs. We demonstrate that indeed, a number of exoplanetary systems hosting planets (or companion brown dwarfs) at large separations from their stars \cite{DeaconEtAl2014,ZhangEtAl2021,RothermichEtAl2024,CifuentesEtAl2025} already provide competitive and independent constraints on the abundance of DMOs in the solar neighborhood. Assuming DMOs originate in Gaussian primordial perturbations, these constraints can be translated into limits on the primordial power spectrum (PPS) itself. Larger amplitudes on scales smaller than those constrained by CMB observations and galaxy surveys generically produce more compact dark halos, which are more prone to dynamically heat their visible environments. Furthermore, in cosmologies with enhanced PPS, the resulting DMO populations can still be rather precisely predicted using standard structure formation theory, enabling robust constraint extraction.

The paper is organized as follows. In \citesec{sec:heating}, we first shortly review the impulsive heating induced by a population of generic DMOs on an exoplanet orbiting its star. Then, in \citesec{sec:constraints}, we discuss our statistical method to extract constraints and the criteria we use to build up the most relevant sample of data for our study. We specialize to UCMHs in \citesec{sec:constraints_ucmhs}, where we derive constraints on the PPS. Finally, we propose new observational signatures beyond constraints in \citesec{sec:signatures}, before concluding.

\section{Heating of exoplanetary systems}
\label{sec:heating}
\begin{figure}
\centering
\includegraphics[width=0.49\textwidth]{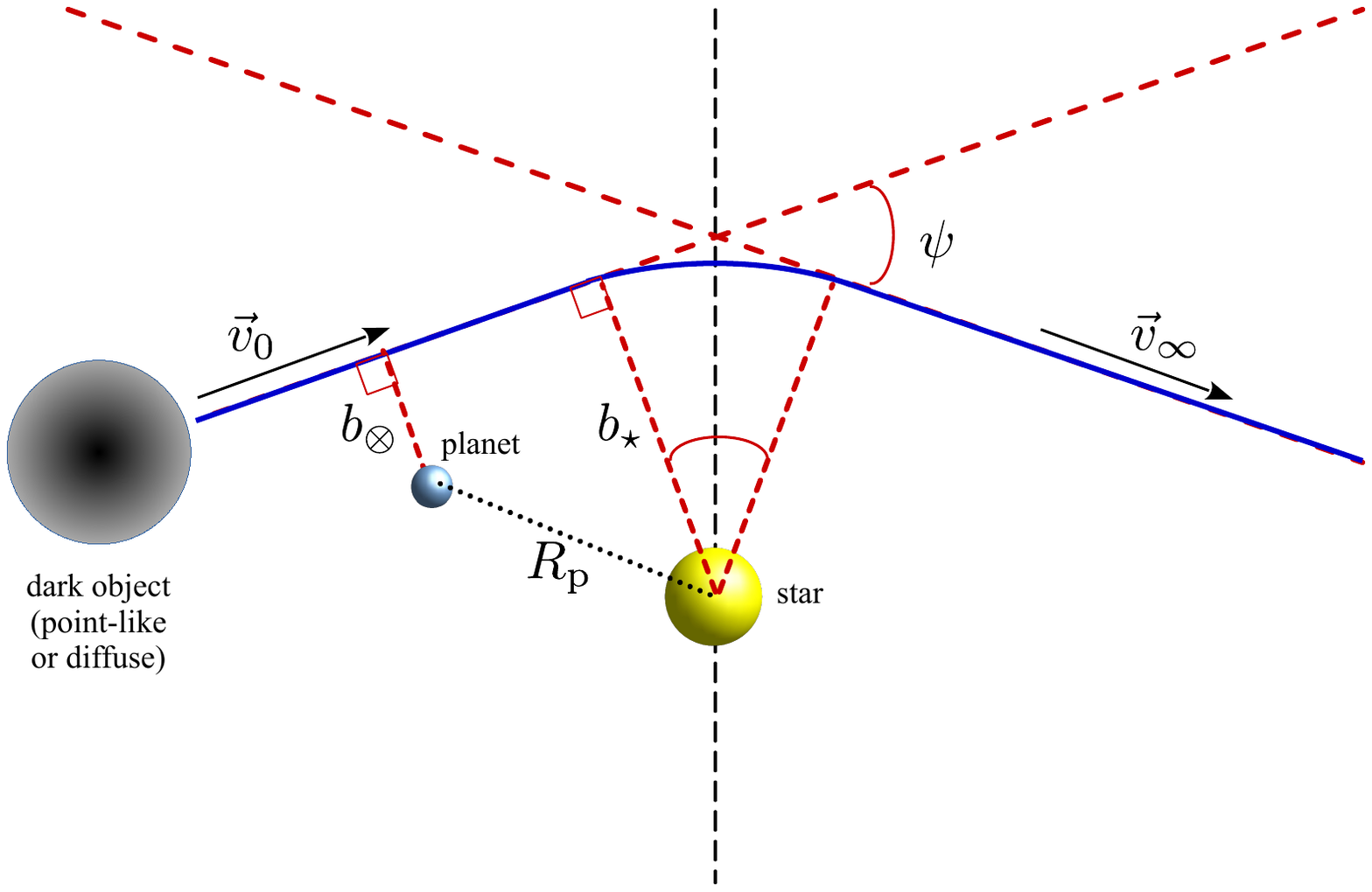}
\caption{\small Hyperbolic encounter between a DMO and a simple exoplanetary system, represented here in a reference frame centered about the star. It would be equivalent to permute the position of the exoplanetary system and that of the DMO.}
\label{fig:encounter}
\end{figure}

The theoretical ground of what follows can be found in textbooks, and the most relevant to this work is very likely the one by Binney and Tremaine~\cite{BinneyEtAl2008}. We adopt a convention such that for any vector $\myvec{x}$, then $x=|\myvec{x}|$.

We consider the simplest possible exoplanetary system: a single planet orbiting its host star on a circular orbit. We assume that the mass of the star is much larger than that of the planet, $\mstar\gg\mplan$, such that the center of mass of the 2-body bound system sits on the star to a good approximation. We further consider the hyperbolic passage of a DMO as exemplified by the drawing of \citefig{fig:encounter}. The change of velocity of the planet in the reference frame of the star induced by such an encounter can be expressed as:
\ben
\dvvplanstar = \dvvplan - \dvvstar
\een
where \dvvplan\ is the velocity kick on the planet and  \dvvstar\ is the velocity kick on the star in an arbitrary Galilean coordinate system. Assuming an impulsive, weak encounter---where the star and planet positions are kept fixed during the encounter, and $|\myvec{\delta v}|=|\myvec{v}_0-\myvec{v}_\infty|\ll |\myvec{v}_0|$ (corresponding to a deflection angle $\psi\approx 0$ in \citefig{fig:encounter})---the relative velocity kick is given by \cite{FacchinettiEtAl2022}:
\ben
\label{eq:dv}
\dvvplanstar = -\dfrac{2G\mdmo}{\vrel}\left[\dfrac{\vbplan}{\bplan^2}I(\bplan)  - \dfrac{\vbstar}{\bstar^2}I(\bstar)\right]\;,
\een
where the relative velocity $\vrel=|\myvec{v}_0|$ is taken the same between the DMO and both the star and the planet ($\vrel=v_0\gg\vplanstar$). In the above equation, $G$ denotes Newton's constant, \mdmo\ the DMO tidal mass, \vbstar\ (\vbplan) the impact parameter to the star (planet), and $I$ an integral accounting for a possible penetrating encounter in the case of an extended DMO, which we will describe below. A useful relation that connects vectors together is the following:
\ben
\label{eq:vecrel}
\vbstar = \vRplan + \vbplan -   \dfrac{\vRplan\cdot\vvrel}{\vrel^2}\,\vvrel\;,
\een
where \vRplan\ is the star-planet vector.

The average heating from a DMO encounter is related to the average change in the planet's kinetic energy per unit mass (in the star's reference frame):
\ben
\label{eq:KE}
\myav{\delta \varepsilon_\planstarsym}&\equiv &\dfrac{\myav{\delta E_\planstarsym}}{\mplan}  =  \vvplanstar\cdot\myav{\dvvplanstar} + \dfrac{\myav{\dvvplanstar^2}}{2}\\
&\simeq & \dfrac{\myav{\dvvplanstar^2}}{2}\;,\nn
\een
where the second equality relies on neglecting internal diffusion and assuming an isotropic distribution of DMOs. The average has to be understood as made over the statistical properties of DMOs. We use the same angular averaging procedure as in \citeref{FacchinettiEtAl2022} (see also \citeref{GerhardEtAl1983}) to obtain:
\ben
\label{eq:dv2}
\myav{\dvvplanstar^2} &=& \(  \dfrac{2G\mdmo}{\myav{\vrel}\bplan}  \)^2\,\Kdmo(\bplan)
= 2\myav{\delta \varepsilon_\planstarsym}\\
\text{with}\;\Kdmo(\bplan) &\equiv&
\( \dfrac{3\bplan^2(I_\plansym-I_\starsym)^2+2\Rplan^2 I_\plansym^2}{3\bplan^2+2\Rplan^2}  \)\;,\nn
\een
where we have introduced a dimensionless function \Kdmo\ of the impact parameter \bplan, defined from the dimensionless integrals $I$ which are generically given for any index $x$ by:
\ben
\label{eq:Iint}
I_x &=& I(b_x) \\
&=& 1 - \dfrac{4\pi}{\mdmo}\Theta(\rdmo-\tilde{b}_x)\int_{\tilde{b}_x}^{\rdmo}\dd x \,
\rhodmo(x)\,x\,\sqrt{x^2-\tilde{b}_x^2}\;.\nn
\een
Here, \rhodmo\ denotes the inner DMO profile, and the DMO mass \mdmo\ and its radial extent \rdmo\ should be understood as tidal quantities that formally depend on position in the Milky Way \cite{FacchinettiEtAl2022}. For $I_\plansym$, we find $\btplan = \bplan$, while for $I_\starsym$, $\btstar = \sqrt{\myav{\vbstar^2}}$, where the angular average reads:
\ben
\myav{\vbstar^2} = \bplan^2+\dfrac{2}{3}\Rplan^2\;.
\een
These $I$ integrals tend to 1 for non-penetrating encounters with DMOs (and are equal to 1 for point-like DMOs), but get suppressed for penetrating ones with a strength that depends on the inner slope of the DMO mass density profile (typically less suppressed for more cuspy profiles). 

A quick order-of-magnitude estimate of the relative kick velocity introduced in \citeeq{eq:dv2} can be obtained by assuming $\Kdmo\sim 1$. By taking an optimal impact parameter $\bplan\sim\Rplan\sim 10^4{\rm a.u.}$, $\vrel\sim 300$~km/s, and $\mdmo\sim 1$~\Msun, we get a significant impulse of $\dvplanstar\sim 59.1$~km/s, with a scaling $\propto \mdmo/\bplan$. Obviously, this has to be balanced by the low probability of such an encounter in the lifetime of the binary system, which we will see later.

The total heating rate per unit mass induced by encounters with a population of DMOs can then be expressed as:
\ben
\label{eq:heat}
\Heatplan &=& \ddfrac{\myav{\Delta \varepsilon_\planstarsym}}{t} =\dfrac{1}{2} \ddfrac{\myav{\Dvvplanstar^2}}{t}\\
&=&
\int\dd\bplan\int \dd \mdmo \ddfrac{^3N_{\dmosym}}{t\,\dd\mdmo\,\dd\bplan} \nn\\
&\times & \acleft \dfrac{\myav{\dvvplanstar^2}}{2} = \myav{\delta \varepsilon_\planstarsym} \acright\,,\nn
\een
where the local differential encounter rate with DMOs is given by:
\ben
\ddfrac{^3 N_{\dmosym}}{t\,\dd \mdmo\,\dd\bplan} = 2\pi \, \bplan \,\myav{\vrel}\,\ddfrac{n_{\dmosym}}{\mdmo}\;.
\label{eq:dNdmodt}
\een
We denote $\myav{\vrel}$ the average relative velocity between the DMOs and the exoplanetary system, which, together with the number density of DMOs in the vicinity of the star $n_{\dmosym}$, helps define the local ``current'' of DMOs. The latter is related to the total density of DM through the fraction \fdmo\ of DM in the form of DMOs in the Galaxy:
\ben
\ddfrac{n_{\dmosym}(\myvec{x})}{\mdmo} = \fdmo\, \dfrac{\rhodm(\myvec{x})}{\myav{\mdmo}}\,\ddfrac{{\cal P}_{\mdmo}(\mdmo,\myvec{x})}{\mdmo} \,,
\label{eq:ndmo}
\een
where $\rhodm(\myvec{x})$ is the Galactic DM halo density at some position $\myvec{x}$, $\myav{\mdmo}$ the average DMO mass over the whole Galactic volume, and $\dd{\cal P}_{\mdmo}(\mdmo,\myvec{x})/\dd \mdmo$ is the DMO mass distribution function, which can in principle be spatial dependent if DMOs are not point-like as a result of gravitational tides in the Milky Way \cite{StrefEtAl2017,FacchinettiEtAl2022}. These tides may lead to the departure of the spatial distribution of DMOs from the smooth Galactic DM halo, especially in the central regions of the Milky Way -- then ${\cal P}_{\mdmo}$ no longer normalizes to unity and may flatten or even suppress the DMO distribution where tides are strong (see the analytic population model described in \citerefs{StrefEtAl2017,FacchinettiEtAl2022}).
%
% Formal definition of the average DMO mass in this model:
% 
%
%

The heating rate given in \citeeq{eq:heat} can therefore be re-expressed at the host star position \vxstar\ as:
\ben
\Heatplan(\vxstar) &=& \dfrac{4\pi G^2}{\myav{\vrel}}
\int\dd \mdmo \,\mdmo^2 \ddfrac{n_{\dmosym}(\vxstar)}{\mdmo}\int_{\bmin}^{\bmax} \dfrac{\dd\bplan}{\bplan}\,\Kdmo(\bplan)\nn\\
&=& \dfrac{4\pi G^2}{\myav{\vrel}}\,\fdmo\dfrac{\rhodm(\vxstar)}{\myav{\mdmo}}
\int\dd \mdmo \,\mdmo^2 \ddfrac{{\cal P}_\dmosym(\mdmo,\vxstar)}{\mdmo}\nn\\
&&\times
\int_{\bmin}^{\bmax} \dfrac{\dd\bplan}{\bplan}\,\Kdmo(\bplan)\;.
\label{eq:heat_full}
\een
At this stage, it is interesting to note that function \Kdmo, defined in \citeeq{eq:dv2}, has two asymptotic regimes set by the star-planet distance scale \Rplan: for encounters with impact parameters to the planet much smaller than the star-planet distance, $\bplan\ll I_{\plansym}\Rplan$, then $\Kdmo\to I_{\plansym}^2$ as if the star were not around, and for distant encounters with $\bplan\gg \Rplan$ (since the $I$'s tend to the same value), then $\Kdmo\to(2/3)(I_{\plansym}\Rplan/\bplan)^2 \propto \bplan^{-2}$. In the above multiple integral, we explicitly set the range of relevant impact parameters. The minimal value, \bmin, is set by requiring a deflection angle less than $\pi/2$ \cite{BinneyEtAl2008}, while the maximal value, \bmax, is set by considering only the largest fluctuation in the gravitational potential due to an impulsive encounter, which amounts to requiring an encounter duration shorter than the exoplanet orbital period $P$ (we consider the time it takes for the potential perturbation to reach its maximum):
\ben
\begin{cases}
  \bmin &= b_{90} = \dfrac{G(\mstar+\mplan+\mdmo)}{\myav{\vrel}^2}\\
  \bmax &= \myav{\vrel}\dfrac{P}{3\sqrt{3}}
\end{cases}\;.
\een
Typical orders of magnitude are $\bmin\approx 4.8$~mpc and $\bmax\approx 59$~pc, for $\vrel\approx 300$~km/s, $\mstar\approx 1$~\Msun, and $\mdmo\approx 10^5\Msun$.
\section{Setting constraints on a population of DMOs}
\label{sec:constraints}
To conservatively constrain DMO populations using exoplanetary systems, we determine the conditions under which tidal heating would disrupt these systems. For a specific planet-star binary system, this amounts to compare the total heating induced by repeating encounters with DMOs during its whole life (age ${\cal T}$) with its binding energy. Disruption typically occurs if the total injected energy
\ben
\label{eq:Rheat}
\varepsilon_{\rm inj}^{\rm tot} &=& \Heatplan\times {\cal T}\geq \dfrac{G\mstar}{2\Rplan}\\
\Rightarrow  \Rheat &\equiv& \frac{2 \varepsilon_{\rm inj}^{\rm tot} \Rplan}{G\mstar}= \dfrac{2\,\Heatplan\, {\cal T}\,\Rplan}{G\mstar}\geq 1\;,\nn
\een
where the ratio \Rheat\ will stand as a diagnosis of disruption, and the heating rate \Heatplan\ is given by \citeeq{eq:heat_full}. A well identified binary exoplanetary system with $\Rheat\geq 1$ should in principle have been disrupted by the heating induced by the DMO population model entering the calculation, which is excluded by the very fact that it is currently observed as a bound system. The challenge in extracting meaningful limits arises because, while averaged values of $\varepsilon_{\rm inj}^{\rm tot}$ are straightforward to compute for DMO models, encounters are stochastic, which introduces statistical fluctuations. We therefore need to assign a probability distribution function (PDF) to the previous inequality. A way to proceed is to link the mean total heat received by an exoplanetary system to an average number of encounters times the average heat received per encounter, as follows:
\ben
\label{eq:etot_av}
\myav{\varepsilon_{\rm inj}^{\rm tot}}(\bar{N}_\dmosym) &=& \bar{N}_\dmosym\myav{\delta \varepsilon_{\planstarsym}}_{(b,\mdmo)}\,,
\een
where we define:
\ben
\label{eq:av_enc}
\bar{N}_\dmosym \equiv \acleft n_{\dmosym}(\vxstar)=\dfrac{\fdmo\rhodm(\vxstar)}{\myav{\mdmo}} \acright\, \pi\,\bmax^2\,\myav{\vrel}\,{\cal T}&&\\
\myav{\delta \varepsilon_{\planstarsym}}_{(\mdmo,b)} \equiv \int\dd \mdmo \ddfrac{{\cal P}_{\dmosym}(\mdmo)}{\mdmo}\(1-\dfrac{\bmin^2}{\bmax^2} \)&&\nn\\
\times
\int_{\bmin}^{\bmax}\dd \bplan \ddfrac{{\cal P}_{\bplan}(\bplan)}{\bplan} \myav{\delta \varepsilon_{\planstarsym}}&&\nn\,.
\een
In the latest definition, the average is performed over both the \mdmo\ mass and impact parameter PDFs, and we further define the latter as: 
\ben
\ddfrac{{\cal P}_{\bplan}(\bplan)}{\bplan} = \frac{2\,\bplan}{\bmax^2-\bmin^2}\,.
\een
Note that this PDF depends on \mdmo\ through \bmin.

In this expression,  $\bar{N}_\dmosym$ can be readily interpreted as the mean number of predicted encounters, assuming a DMO population model. A quick order-of-magnitude estimate of the encounter rate can be obtained by calculating the ratio $\bar \Gamma_\dmosym\equiv \bar{N}_\dmosym/{\cal T}$, which is independent of the age ${\cal T}$ of the binary system, as it should. Assuming point-like DMOs of $\mdmo\sim 10^5$~\Msun, a fraction $\fdmo\sim 0.1$, a local DM density $\brhodm\sim 0.01\,\Msun/{\rm pc}^3$, $\vrel\sim 300$~km/s, and a maximum impact parameter tuned to an optimal value of $\bmax\sim\Rplan\sim 10^4{\rm a.u.}$, we get $\bar \Gamma_\dmosym\sim 2.3\times 10^{-5}{\rm Gyr}^{-1}$. This confirms that close encounters should be rare, and that stochasticity matters a lot in the heating process we are considering.

Although numerically possible in principle, it turns rather cumbersome to project the mass and impact parameter distributions onto a probability distribution for the total heat energy. Instead, assigning a Poisson distribution probability of mean $\bar{N}_\dmosym$ to the number of encounters is an alternative way to extract a statistical meaning to the total injected energy of the form:
\ben
\label{eq:poisson}
\prob(\varepsilon_{\rm inj}^{\rm tot}= N_\dmosym \myav{\delta \varepsilon_{\planstarsym}}_{(b,\mdmo)})=\prob_{\rm pois}(N_\dmosym|\bar{N}_\dmosym)\,.
\een
The dependence on the DMO properties enters both the expressions of $\bar{N}_\dmosym$  and $\myav{\varepsilon_{\rm inj}}_{(\mdmo,b)}$; the DM fraction in DMOs, \fdmo, only enters the expression of $\bar{N}_\dmosym$. These properties can either be set by hand or predicted from a model. We can further define the probability of tidal disruption for a single system as:
\ben
\label{eq:pdisr}
\prob_{\rm disr} = \prob(N_\dmosym^{\rm disr}|\bar{N}_\dmosym) = \prob({\cal R}_{\plansym}^{\rm heat} > 1 | \bRheat )\;,
\een
where the average heating-to-binding energy ratio \bRheat\ is obtained by injecting the average total heating energy of \citeeq{eq:etot_av} into  \citeeq{eq:Rheat}. $\prob_{\rm disr}$ is the probability of having the number $N_\dmosym^{\rm disr}$ of encounters required to heat the system up to disruption (equivalently, ${\cal R}_{\plansym}^{\rm heat} > 1$), given the predicted values for $\bar{N}_\dmosym$ and $ \myav{\delta \varepsilon_{\planstarsym}}_{(b,\mdmo)}$ introduced in \citeeq{eq:av_enc} (equivalently \bRheat), which depend on the DMO model parameters.

For compact DMOs, we can derive a rough estimate on the local fraction of them \fdmo\ we can constrain. For ``light'' DMOs, typically lighter than $\sim 10^{6}\Msun$, we are in the regime where the impact parameters can be of order or less than \Rplan, and we get:
\ben
%\label{eq:est_flight}
\fdmo&\lesssim& 4\ttpow{-4} \(\dfrac{\brhodm}{0.01\,\Msun/{\rm pc}^3} \)^{-1}\(\dfrac{\mstar}{\Msun} \)^{-1}% \\
%&&\times
\(\dfrac{\mdmo}{\tpow{6}\Msun} \) \nn\,.
\\
\een
In the ``heavy'' regime, we get instead:
\ben
\fdmo &\lesssim& 6\ttpow{-3} \(\dfrac{\brhodm}{0.01\,\Msun/{\rm pc}^3} \)^{-1}\(\dfrac{\mstar}{\Msun}\)\(\dfrac{\mdmo}{10^6 \Msun}\)\nn\\
&&\times \(\dfrac{\Rplan}{10^4{\rm a.u.}}\)^{-3}\(\dfrac{\cal T}{5\,{\rm Gyr}}\)^{-1}\(\frac{\vrel}{300\,{\rm km/s}}\)^{-3}\,.
\label{eq:est_flight}
\een
%% HERE HERE JL
%
%
\section{Exoplanetary constraints on the primordial power spectrum}
\label{sec:constraints_ucmhs}
\begin{figure}
    \centering
    \includegraphics[trim={3cm 0 3cm 0},clip, width=0.49\textwidth]{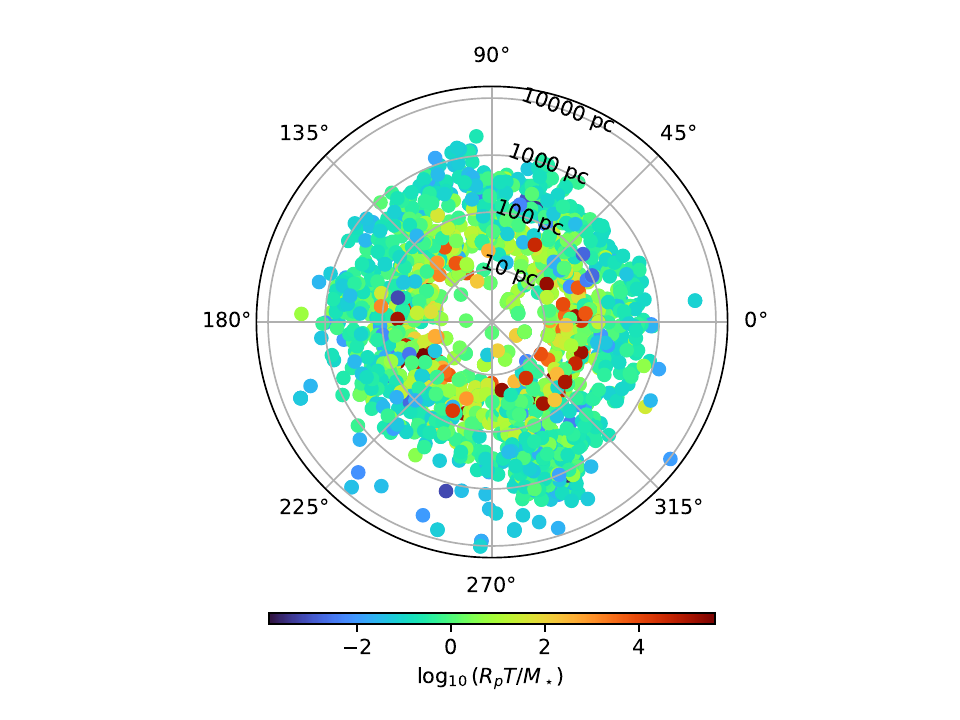}
    \caption{\small Distribution of exoplanets by distance to the Sun and right ascension, color-coded by their constraining power (see text for details). Figure generated from the data collected from \href{https://exoplanet.eu/catalog/}{https://exoplanet.eu/catalog/}.}
    \label{fig:planetpos}
\end{figure}
We now specialize to a peculiar class of DMOs: those arising from a modification to the standard scale-invariant primordial power spectrum, where a peak is injected at a given scale, which can occur in non-minimal models of inflation (see \eg~\citerefs{Starobinskii1992,BallesterosEtAl2018a,ByrnesEtAl2019}). This is prototypical of ways to get simple scenarios of almost monochromatic PBH populations, except that by limiting the amplitude of the peak, we can actually enter a regime where the production of PBHs is suppressed (critical fluctuations are too rare) to the benefit of ultra-compact minihalos (UCMHs) --- this corresponds to density fluctuations of  $10^{-3}\lesssim \delta \lesssim 10^{-1}$ which enter the horizon deep in radiation domination and may collapse before matter-radiation equality (see \eg~\citerefs{BerezinskyEtAl2003,BerezinskyEtAl2010,BerezinskyEtAl2011,BerezinskyEtAl2014,RicottiEtAl2009,DelosEtAl2018a,DelosEtAl2023}). Here, we follow \citerefs{AbellanEtAl2023,DelosEtAl2018a} to derive a combined cosmological mass function for both UCMHs and standard CDM halos forming out of perturbations arising from a primordial power spectrum enhanced by a peak of large amplitude  population model. All this derives from the following assumption for the (here dimensionless) primordial power spectrum:
\ben
\label{eq:pps}
%{\mathdutchcal P}(k) &=& {\cal P}_{\rm pl}(k)+{\cal P}_\dmosym(k)\\
\dlpspecR(k) &=& \dlpspec_\text{p-l}(k)+\dlpspec_\dmosym(k)\\
&=&  {\cal A}_s\(\frac{k}{k_s}\)^{n_s-1} +
{\cal A}_{\dmosym}\kdmo\delta \(k-\kdmo \) \,\nn
\een
where $k$ denotes the comoving wavenumber (our \dlpspecR\ can also be termed $\Delta_{\cal R}^2$ in the literature). This corresponds to the standard power-law primordial power spectrum $\dlpspec_\text{p-l}$ augmented by an additional peak term $\dlpspec_\dmosym$ that features two free parameters, the peak injection scale \kdmo\ and the corresponding amplitude ${\cal A}_{\dmosym}$. Parameters ${\cal A}_s=2.1\ttpow{-9}$, $k_s=0.05$~Mpc$^{-1}$, and $n_s=0.96$ are fixed to their constrained Planck-2018 values \cite{PlanckCollab2020}. From this primordial power spectrum, we can determine the matter power spectrum, then the cosmological mass function. The latter is the building piece of information to derive a theoretically consistent UCMH population model based on standard structure formation theory, in the framework of the excursion set approach \cite{PressEtAl1974,BondEtAl1991a,LaceyEtAl1993}. This allows us to fully {\em predict} the abundance, mass function, and spatial distribution of DMOs in host halos of any mass and at any redshift. Other properties can also be included, such as the baryon distribution, if the population model is applied to a constrained observed host galaxy like the Milky Way. More details are given in \citeapp{app:ucmhs}, and an illustration of the associated DM subhalo mass function for a Milky Way-like galaxy is showed in \citefig{fig:UCMH_mf}, which consistently includes a subpopulation of UCMHs.

\begin{figure}[t]
\centering
\includegraphics[width=0.49\textwidth]{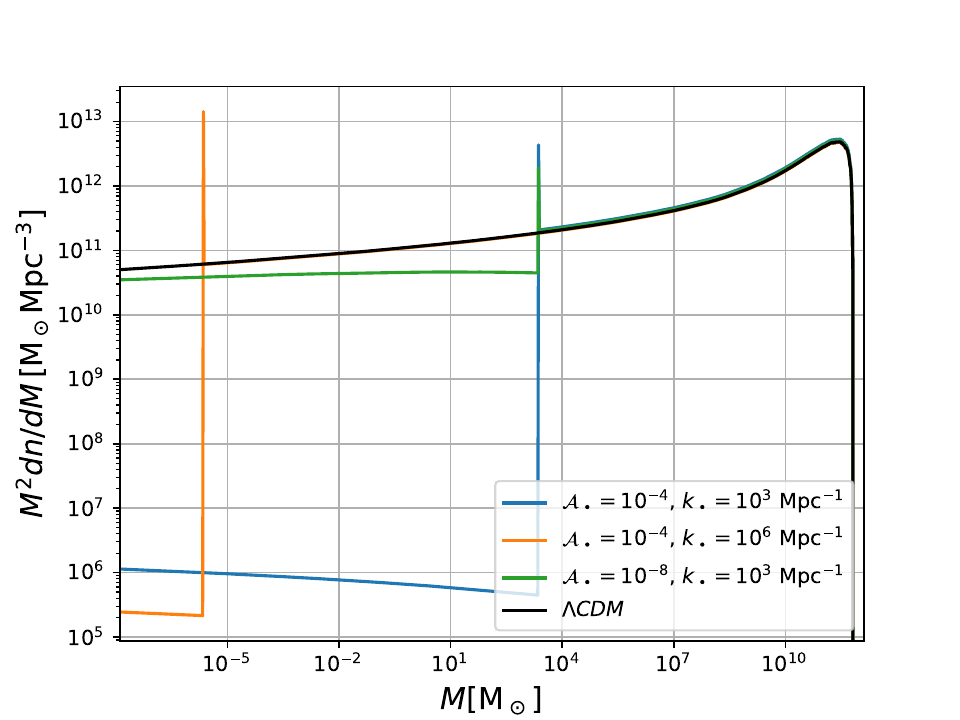}
\caption{Prediction of the DM subhalo mass function for a Milky Way-like galaxy at redshift $z=0$, including UCMHs for different peak amplitudes and injection scales.}
\label{fig:UCMH_mf}
\end{figure}

\begin{figure*}[t]
\centering
\includegraphics[width=0.49\textwidth]{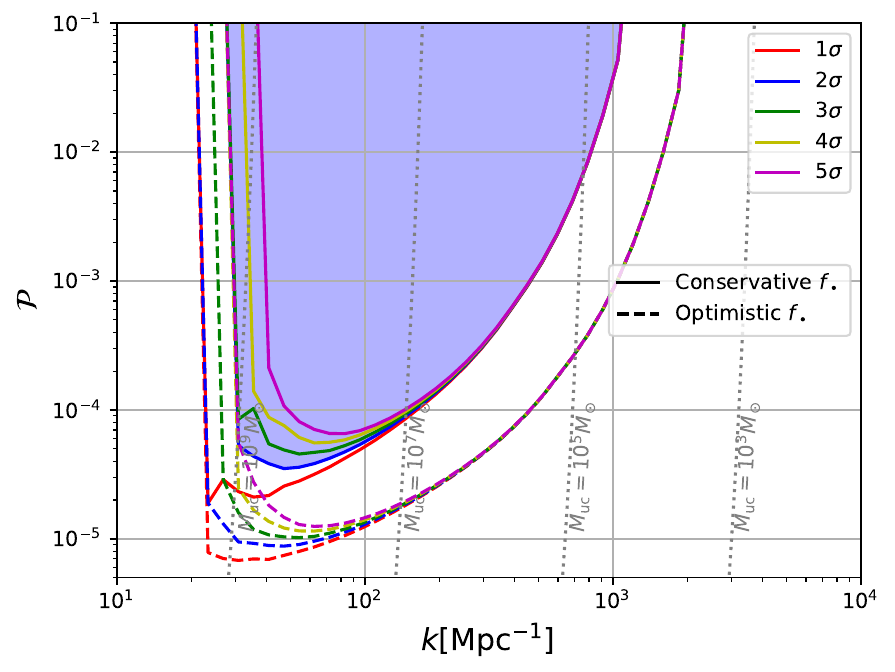}
\includegraphics[width=0.49\textwidth]{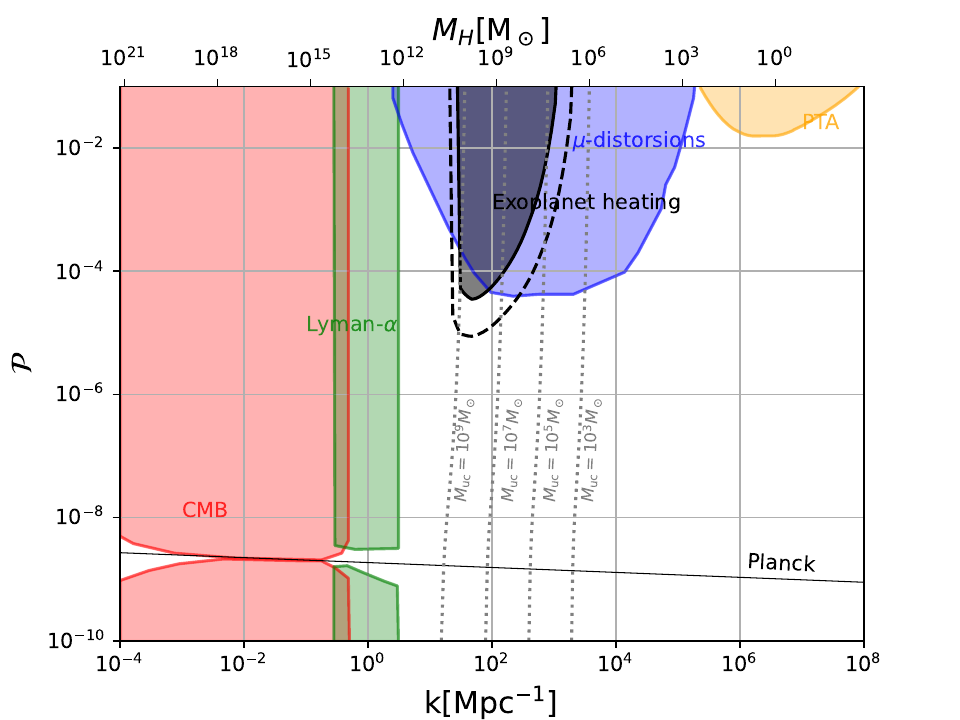}
\caption{\small {\bf Left panel}: Constraints on the primordial power spectrum from the disruption of exoplanetary systems. {\bf Right panel} Comparison with other constraints on the primordial power spectrum --- see the text for details.}
\label{fig:limits}
\end{figure*}

We can constrain Galactic UCMH populations generated by varying \kdmo\ and ${\cal A}_{\dmosym}$ thanks to exoplanetary systems with low binding energies, for which the injected-to-binding energy ratio ${{\cal R}_{\plansym}^{\rm heat}}$ given in \citeeq{eq:Rheat} is large. We actually predict two asymptotic regimes for this ratio, ${{\cal R}_{\plansym}^{\rm heat}}\propto \Rplan {\cal T}/\mstar$ for light UCMHs with $\bmin\ll\Rplan$, and $\propto \Rplan^3 {\cal T}/\mstar$ for heavy ones with $\bmin\gg\Rplan$, which helps select our exoplanet sample. We found that demanding $\log_{10}(\Rplan {\cal T}/\mstar)>2$ or $\log_{10}(\Rplan^3 {\cal T}/\mstar)>5$, where \Rplan, ${\cal T}$, and \mstar\ are expressed in units of a.u., Gyr, and \Msun, respectively, was a sufficient criterion. A position distribution (with respect to us) of currently observed exoplanets together with one of these selection criteria is shown in \citefig{fig:planetpos}. We also provide in \citetab{tab:exoplanets} the list of the dozen exoplanets we found the most constraining in our analysis, out of a selected sample of 128 objects. Several of them were found rather recently (see~\eg~\citeref{RothermichEtAl2024}). We performed our data selection from the catalog provided by the Exoplanet Encyclopaedia exoplanet.eu. We give more details about the characteristics and relevance of our data sample in a dedicated appendix section, \citeapp{app:data}.
\begin{table}
\centering
\begin{tabular}{ c| c |c }
 Name & $\log_{10}(\Rplan T/\mstar)$ & $\log_{10}(\Rplan^3T/\mstar)$ \\ 
 \hline
 NLTT 35024 c & 5.63 & 14.39 \\  
 G 194-47 (AB)b &5.48 & 12.99   \\
 CWISE J1332-3749 b &5.44 & 12.92\\
 CWISE J1015-1115 (AB)b  &5.37 & 12.90\\
 HD 188769 (AB)b  &5.37 & 14.17\\
 2MASS J0953-5055 b  & 5.35& 12.00\\
 BD+24 4329 b  & 5.34& 14.50\\
 HD 170573 b &5.31 & 13.46\\
 Ross 19 b  & 5.29& 13.29\\
 LEHPM 5083 b  &5.22 & 12.40\\
 WD J2122+6600 b &5.06 & 13.96\\
 UCAC4 840-013771 b &4.26 & 12.81
\end{tabular}
\caption{\small List of the 12 most constraining exoplanetary systems. It contains the 10 most constraining planets in both heating regimes referred to in the text with 8 planets common to the two regimes.}
\label{tab:exoplanets}
\end{table}

We combined all of these exoplanets together to strengthen the limits on UCMHs by defining a global survival probability at $n\sigma$ confidence level (CL, in the Gaussian sense):
\ben
%{\cal P}_\mathrm{surv}^{(n\sigma)} =\prod_{i\in \{\mathrm{planets}\}}
{\cal P}_\mathrm{surv} =\prod_{i\in \{\mathrm{planets}\}}
 \left[1-{\cal P}_{\rm disr}(N_i^{\rm disr}|\bar N_{\dmosym,i})\right]\,,
\een
where, for simplicity, we consider the individual survival probabilities to be independent from each other. They derive from the individual disruption probabilities given by ${\cal P}_{\rm disr}(N_i^{\rm disr}|\bar N_{\dmosym,i})$ given in \citeeq{eq:pdisr}. We can therefore extract limits at $n\sigma$ CL (in the Gaussian statistical sense) by requiring ${\cal P}_\mathrm{surv}^{(n\sigma)}\geq n\sigma$.

We show our limits on the primordial power spectrum in \citefig{fig:limits}, where we make explicit the reach as a function of the CL, from 1 to 5$\sigma$, in the left panel. In this panel, we show the results corresponding to two different assumptions for the Galactic UCMH population: (i) very conservatively, we do not count those UCMHs which have been accreted into bigger subhalos before merging into the Galactic halo during its formation history (solid curves); (ii) we count them in the UCMH budget, assuming they are so dense that they have not diluted in their pre-merging subhalos (dashed curves). In the right panel, we compare these limits to others extracted from other probes: CMB anisotropies \cite{PlanckCollab2020}, the Lyman-$\alpha$ forest \cite{BirdEtAl2011}, CMB spectral distortions \cite{ChlubaEtAl2012a,KohriEtAl2014}\footnote{See also \citeref{CroonEtAl2024} for CMB constraints in terms of fraction of DM in extended DM objects, not connected to specific properties of primordial perturbations, and with different inner density profiles.}, pulsar timing arrays \cite{ByrnesEtAl2019,InomataEtAl2019}. On the two panels, we explicitly report as dotted (almost) vertical dotted lines reference UCMH masses which correspond to the two parameters of our model, \kdmo\ and ${\cal A}_\dmosym$. We see that the exoplanetary probe is essentially sensitive to UCMHs more massive than $\sim 10^5\msun$.

\section{Possible signatures beside disruption}
\label{sec:signatures}

\begin{figure*}
\centering
\includegraphics[width=0.99\textwidth]{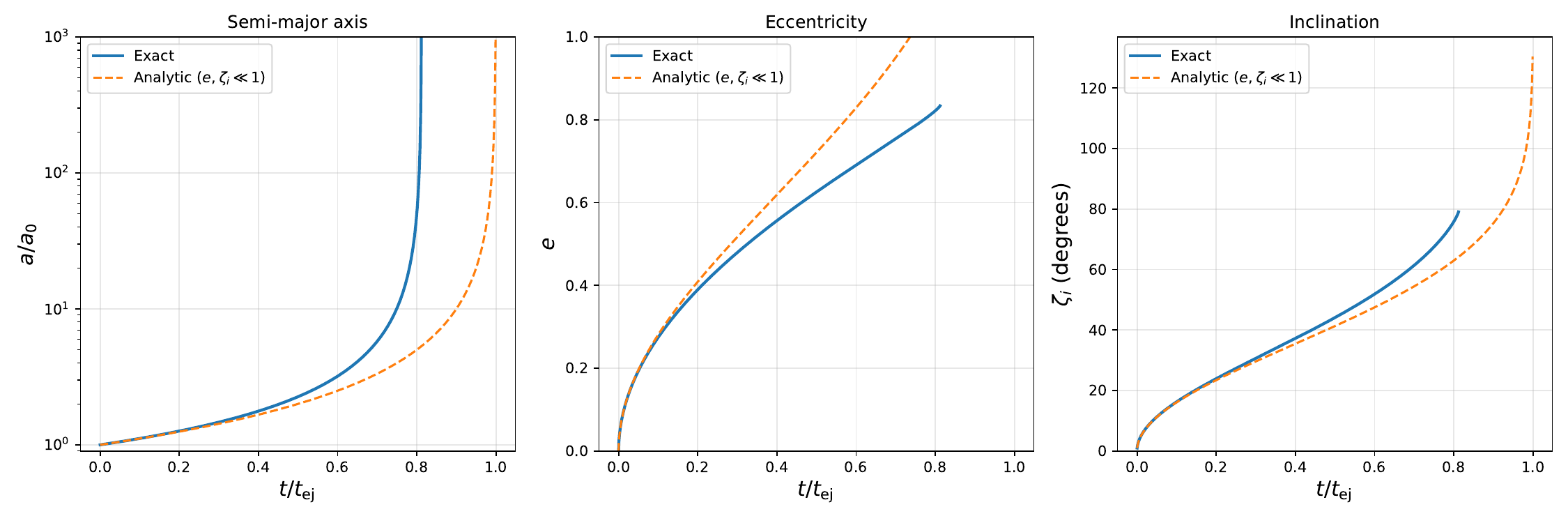}
\caption{\small {\bf Left panel:} Evolution of the semi-major axis as a function of time for a constant heating rate. The low-eccentricity approximation underestimates the heating and the actual ejection time is actually about $80\%$ of the analytical value $\tej$. {\bf Middle panel:} Corresponding evolution of the eccentricity. The exact solution reaches an asymptotic value where $\dot{(e^2)}\to 0$. {\bf Right panel:} Corresponding evolution of the inclination. The analytical approximation underestimates the heating, the final value at ejection is $53.5^\circ$.}
\label{fig:migration}
\end{figure*}

In this section, we use the adiabatic invariants of orbital motion--the actions--whose evolution in time can be directly related to impulsive heating. These are classical tools in stellar dynamics, and we refer to \citerefs{BinneyEtAl2008,Binney2026} for technical details.

All the proofs of what follows can be found in a dedicated appendix section, \citeapp{app:orbits}, of which we only summarize the main results for clarity. The question we tackle here is whether we can relate the heating rate, given a DMO model, to the time evolution of the main orbital parameters, which we choose to be the semi-major axis $a$, the eccentricity $e$ (we will actually consider the squared eccentricity $e^2$), and the inclination angle $\zeta_i$ between the initial orbital plane and the time of observation --- the latter can only make sense for multiplanetary systems, where the initial orbital plane could then be identified from the other planets orbiting closer to their stars.

We first consider the radial action, which can be written as:
\ben
J_r = \frac{G\mstar}{\sqrt{2\mE}} - L\nn\,,
\een
where the total orbital energy \mE\ (here positive defined for a bound orbit) and the total angular momentum $L$ are given by:
\ben
%\label{eq:ELexc}
\mE = \frac{G\mstar}{2\,a}\quad, \quad L = \sqrt{G\mstar a(1-e^2)}\,.
\een
We further consider the vertical action, which can be expressed as:
\ben
J_z = L-|L_z| = L(1-|\cos\zeta_i|)\,.
\een
Actions $J_r$ and $J_z$ are adiabatic invariants whose time evolution can be directly related to impulsive heating as:
\ben
\ddfrac{J_r}{t} = \ddfrac{J_z}{t} = \dfrac{P}{2\pi}\Heatplan\,.
\een
By further connecting the time evolution of the angular momentum to the heating as:
\ben
\ddfrac{L}{t} = \dfrac{a^2\Heatplan}{3L} \left( 1+\dfrac{3}{2}e^2\right)\,,
\een
we obtain the following closed system of time differential equations for the orbital parameters:
\ben
\label{eq:orbit_evol}
\begin{cases}
\dfrac{\dot a}{a} &= \dfrac{2\,a\,\Heatplan}{G\mstar} \left(1 + \dfrac{2+3\,e^2}{6\sqrt{1-e^2}}\right)\\
\dot{(e^2)} &= \dfrac{2\, a\, \Heatplan}{G\mstar} \left[\dfrac{2}{3} - \dfrac{3e^2}{2} + \left(\dfrac{1}{3} + \dfrac{e^2}{2} \right)\sqrt{1-e^2}\right]\\
\dot \zeta_i\sin \zeta_i &= \dfrac{a\,\Heatplan}{G\mstar}\left[\dfrac{1}{\sqrt{1-e^2}} - \dfrac{2+3e^2}{6(1-e^2)}(1-\cos \zeta_i)\right]
\end{cases}\,.
\een
This system can be solved numerically. In the limit of low eccentricity, we can get the following approximate analytical solutions:
\ben
\begin{cases}
a(t) &= \dfrac{a_0}{1 - t/\tej}\;,\;
\text{with}\;\tej \equiv \dfrac{3G\mstar}{8a_0\Heatplan}\\
e^2(t) &= e_0^2 + \dfrac{3}{4} \ln \left( \dfrac{a(t)}{a_0} \right)\\
\zeta_i^2(t) &= \zeta_{i,0}^2 + \dfrac{3}{4}\ln\left(\dfrac{a(t)}{a_0}\right)
\end{cases}\,.
\een
These approximate analytical solutions are depicted in \citefig{fig:migration} (dashed curves), where we see that they match reasonably well with the exact numerical solutions to the system of \citeeq{eq:orbit_evol}. These analytical approximations actually underestimate the change in semi-major axis, while they overestimate the change in eccentricity. What is particularly striking is that the semi-major axis does not evolve so much before the planet gets rapidly ejected around \tej. That means that impulsive heating is not likely to have pushed the planet from an inner orbit away to a very distant orbit in the first place - consequently, extracting limits on such external heating from evaporation or disruption makes perfectly sense.

All of the results illustrated in \citefig{fig:migration} allow us to formulate potential signatures specific to the impulsive heating induced by external invisible perturbers like DMOs: the progressive misalignment of distant planets as function of their distance to their star, with respect to the main orbital plane characterized by planets closer to their star. Observing such properties would require data from multiple systems hosting both very distant ($\sim 10^4$ a.u.) and very close-in planets. This goes beyond the scope of this work, but would be an interesting test to make on the long run, as the amount of data is expected to increase significantly in the near future.

%At second order in $\zeta_i$ and $e$ we can derive:
%\ben
%\dot{(\zeta_i^2)} = \dfrac{2\,a\,\Heatplan}{G\mstar}\left(1 + \dfrac{e^2}{2} -\dfrac{\zeta_i^2}{6}\right)\,,
%\een
%where we see that the correction is increasing the efficiency of the heating as the inclination and the eccentricity grow larger.

%\clearpage
\section{Conclusion}
\label{sec:concl}
In this work, we have proposed ultra-wide-orbit exoplanets as a novel probe of Galactic populations of moderately compact dark matter objects. These objects may arise from large primordial perturbations, intermediate between the regimes of primordial black holes and of standard cold dark matter subhalos. Such large perturbations can be produced from a variety of phenomena in the early universe, for instance inflation beyond its canonical description or phase transitions (etc.), and generically lead to the formation of dark ultra-compact minihalos, which are expected to populate galaxies in numbers. These dark populations may dynamically heat visible systems, more specifically wide binary systems. Ultra-wide-orbit exoplanets (or companion brown dwarfs) at large separations from their host stars, typically detected from direct imaging, are ideal candidates for this purpose. Compared to wide stellar binaries (previously used to search for dark objects \cite{PenarrubiaEtAl2010a,PenarrubiaEtAl2016,RamirezEtAl2023a,TylerEtAl2023,ShariatEtAl2025,BhallaEtAl2025}), exoplanetary systems offer a key advantage: their ``binarity'' can in principle be aged rather precisely --- individual stars in stellar binaries can also be aged rather precisely too, but the time they spent in the form of a binary system is more difficult to estimate (we could even push the argument further and cook up a scenario in which some of them could form from encounters with dark objects in the first place). Aging the ``binarity'' as precisely as possible is critical as heating is proportional to age.

We began by introducing the general framework for impulsive heating in \citesec{sec:heating}, followed by our statistical methodology to set limits in \citesec{sec:constraints}. Then, in \citesec{sec:constraints_ucmhs}, we specialized to ultra-compact minihalos, where a key step was predicting the properties of their Galactic population. This endeavor is essential to properly connect constraints on DMO masses and fraction to the fundamental parameters characterizing their primordial origin. This step, detailed in \citeapp{app:ucmhs}, represents a significant effort in this study. By gathering a sample of old and wide-orbit exoplanetary systems, we were able to derive limits competitive with those from cosmic microwave background and pulsar-timing-array data. These results, illustrated in \citefig{fig:limits} for the case of Gaussian perturbations, open up a promising and complementary avenue to probe the fate of large primordial perturbations.

Beyond constraints, we also proposed in \citesec{sec:signatures} new observational signatures of dark matter objects transiting around exoplanetary systems. These includes the progressive inclination or misalignment of orbital planes with increasing semi-major axes in multiplanetary systems. While such signatures require the observation of systems with both wide and close orbits, they would provide an exquisite test of dark populations beyond tidal disruption limits.

These results pave the way for future development and improvement, and strengthen the case for exoplanetary science \cite{DeegEtAl2026} beyond the astrophysics of stars and planets only, with additional items on properties of the primordial universe. As we saw, ultra-wide-orbit exoplanets are typically detected via direct imaging techniques, and upcoming observations with for instance the Nancy Roman Space Telescope or next-generation extremely large ground-based telescopes offer bright perspectives to yield even more data relevant to Galactic dark object searches \cite{SpergelEtAl2015,Chauvin2023}. More broadly, this probe complements existing gravitational searches for these exotic tracers of both primordial perturbations on small scales and dark matter, whether through dynamical heating \cite{LaceyEtAl1985,JohnstonEtAl2002,PenarrubiaEtAl2016,Penarrubia2018,ChiangEtAl2023,GrahamEtAl2026}, lensing, or timing signatures \cite{DrorEtAl2019,GilmanEtAl2026,BringmannEtAl2025,Ando2026}.

\begin{acknowledgments}
We thank the regular participants in the {\em News from the Dark} workshop series for the stimulating exchanges that accompanied this research project, in particular James Binney, Raphael Errani, and Benoit Famaey. We also thank M.~Sten Delos for useful discussions. This work has been supported by funding from the ANR project ANR-18-CE31-0006 ({\em GaDaMa}), and from European Union's Horizon 2020 research and innovation program under the Marie Sk\l{}odowska-Curie grant agreement N$^{\rm o}$ 101086085–ASYMMETRY. We also benefited from recurrent public funding from CNRS and the University of Montpellier. TP is supported by a PhD excellence grant from CNRS-IN2P3. This research has made use of data obtained from or tools provided by the portal exoplanet.eu of The Extrasolar Planets Encyclopaedia.
\end{acknowledgments}

\newpage
\appendix

\section{Galactic UCMH population model}
\label{app:ucmhs}
In this appendix section, we describe how we predict the abundance and mass function of a population of UCMHs in a Milky Way-like galaxy at redshift $z=0$, using standard theory of structure formation, \eg~\citeref{MoEtAl2010}. At the cosmological level, we will follow the approach developed in \citeref{AbellanEtAl2023}, and will resort to a semi-analytic merger-tree approach \cite{LaceyEtAl1993,ColeEtAl2000,ParkinsonEtAl2008} at the host galaxy level. As already implicit in the expression of the primordial power spectrum given in \citeeq{eq:pps}, any host galaxy will be populated by two different classes of objects: (i) standard DM subhalos coming from the nominal power-law part of the power spectrum, and (ii) UCMHs originating from the peak part. An interesting feature of any prediction of abundance or mass function, be it at the cosmological level or at a host halo level even on small scales, is that it relies only on the linear theory, in contrast with angular power spectra. The non-linear character of structure formation is all contained in the critical threshold for a density contrast to collapse. This explains why probing or constraining DMOs on small scales in a single host galaxy can give powerful information, invertible in terms of cosmological parameters.

We first describe the classic procedure to generate a standard DM merger tree, before complementing the discussion with UCMHs. We note that many aspects of what follows have been already used in other studies and are at the heart of existing semi-analytic codes, \eg~\citerefs{Benson2012,JiangEtAl2020,HiroshimaEtAl2022}. It builds upon the excursion-set theory (EST) \cite{BondEtAl1991a}, also called extended Press-Schechter (EPS) theory after \citeref{PressEtAl1974} --- see \eg~\citeref{Zentner2007} for a review. We assume that a halo can form when the local dimensionless density contrast over some spatial scale $R$ exceeds a fixed threshold \deltac\ (all scales smaller than $R$ are smeared out). We give a pedagogical and self-contained summary of the overall procedure below.

\subsection{The standard lore}
To go from the matter density contrast field $\delta$ to the matter density field $\delta_R$ averaged over some comoving scale $R$, we resort to a filtering window function in real comoving space, $W_R$ --- for the moment, we forget about the redshift or time dependence on purpose (it factors out, and will be recovered later on); all quantities are evaluated at $z=0$, unless specified otherwise. This window function is normalized to unity in the whole comoving volume, such that the filtered contrast can be defined as:
\ben
\label{eq:deltaR}
\delta_R(\vec{r}) \equiv \int\dd^3 \vec{r'}\, \delta(\vec{r'}) \,W_R(|\vec{r}-\vec{r'}|)\;.
\een
In Fourier space, this becomes
\ben
\delta_k(R) = \hat{W}_R(k)\,\delta_k\,,
\een
where $\hat{W}_R(k)$ and $\delta_k$ are the Fourier transforms of $W_R$ and $\delta(\vec{r})$, respectively. Following \citeref{AbellanEtAl2023}, here we adopt the $k$-space top-hat window function in Fourier space, which reads: 
\beq
\hat W_R(k) = \hat W_R^{k\text{-th}}(k) = \Theta(1-kR)\,.
\eeq
It is particularly relevant for a Gaussian density field, which we assume here. The corresponding would-be halo mass $M$ is defined from an effective volume $V(R)$ such that, by convention,
\ben
M = \rhomattO\,V(R)\,,
\een
where $\rhomattO$ and $R$ are the matter density and the comoving filtering scale today, respectively. The effective volume $V(R)$ should in principle relate to the normalization condition for $W_R$ in real space, but the latter actually diverges for the $k$-top-hat case. The conventional alternative definition is $V(R) = V_{k\text{-th}}= 6\pi^2R^3$, which we adopt here.

All this allows us to extract the variance of the density field filtered on a comoving scale $R$ as its autocorrelation:
\ben
\label{eq:defS}
S(R)\equiv \sigma^2(R) &=& \myav{\delta_R(0)\delta_R(0)} \\
&=& \int_0^\infty\dd \ln k\, \Dmpspec^2(k)\,|\hat W_R(k)|^2\nn
\een
where we have introduced the dimensionless matter power spectrum
\ben
\Delta^2_{\rm m} (k)\equiv \frac{k^3}{2\pi^2}\mpspec(k)\,,
\een
expressed in terms of the matter power spectrum \mpspec, evaluated at $z=0$. As a side detail, note that in \citeeq{eq:pps} we have used a different notation for the dimensionless formulation of the primordial power spectrum, which may vary depending on references, though still related to its dimensionful analog as follows:
\ben
\Delta^2_\mathcal{R}(k) \equiv \dlpspecR(k) = \frac{k^3}{2\pi^2}\pspecR(k)\,.
\een
The matter power spectrum relates to the primordial power spectrum \pspecR\ by means of the following equation \cite{Dodelson2003}:
\ben
\label{eq:mpspec}
\mpspec(k,z) &\equiv& \dfrac{4}{25}\[\dfrac{T_{\rm tr}(k)\,D(z)\,k^2}{\OmattO H_0^2} \]^2\,  \pspecR(k)\\
\mpspec(k) &\equiv& \mpspec(k,z=0)\nn\;,
\een
where $T_{\rm tr}(k)$ is the transfer function that encodes the scale dependence of the density contrast, and which suppresses modes that entered the horizon before matter domination, $\Omatt^0$ and $H_0$ stand for the present values of the relative matter energy density and Hubble expansion rate, respectively. Note that it is here that we temporally reintroduce the time or redshift dependence through $D(z)$, the growth factor, which encodes the growth of perturbations thanks to gravity and will turn important later on. Therefore, the variance $S(R)$ has to be understood as a time-dependent function $S(R,z)\propto D^2(z) $ on general grounds, even though it is evaluated at $z=0$ in the current discussion. We take the following definition for the growth factor \cite{Heath1977,CarrollEtAl1992,MoEtAl2010}:
\ben
D(z) &\equiv & \dfrac{d(z)}{d(z=0)}\\
\text{with}\;d(z)&=&d(a=(1+z)^{-1})\nn\\
&\equiv& \dfrac{H(a)}{H_0}\int_{\sim 0}^{a}\dd a'\,\(\dfrac{H_0}{a'\,H(a')}\)^3\,,
%\dfrac{5\,\Omatt(z)\,g^{-1}(z)}{2(1+z)}\\
%\text{with}\;g(z) &\equiv & \Omatt^{4/7}(z) - \OLambda(z)\nn\\
%&&+ \[1+\dfrac{\Omatt(z)}{2} \]\[1+\dfrac{\OLambda(z)}{70} \]\,,\nn
\een
which we tabulate as a function of redshift or scale factor for a given set of cosmological parameters --- $D(z=0)=1$ from this definition. For the transfer function $T_{\rm tr}(k)$ featuring \citeeq{eq:mpspec}, we adopt the parameterization given in \citeref{EisensteinEtAl1998}.

In EST, one exploits the direct and monotonous mapping between the variance $S(R)$ of the smeared density field and the scale $R$ (or mass scale $M$) at each slice of time or redshift, with $S$ growing from 0 to infinity as $1/R$ does so. The number of collapsed halos is related to the number of $\delta_R$ trajectories crossing a threshold $\deltac$, which we set to 1.686 here. Starting at a given space-time point of contrast $\delta_0$ and decreasing the spatial/mass scale from a very small initial variance $S_0$ (corresponding to an infinitely large spatial scale $R$), the density of trajectories having a density contrast $\delta$ at variance $S=S(R)$ (\ie~at some finite scale $R$) reads:
\ben
&& p(\delta,S) = \dfrac{1}{\sqrt{2\pi\Delta S}}\\
&&\times \acleft \exp\( -\dfrac{(\Delta \delta)^2}{2\Delta S}\)
-  \exp\( -\dfrac{(2\Delta\deltac-\Delta\delta)^2}{2\Delta S}\) \acright\;.\nn
\een
where $\Delta\delta\equiv(\delta-\delta_0)$, $\Delta\deltac\equiv \deltac-\delta_0$, and $\Delta S\equiv (S-S_0)$. The fraction of collapsed objects of masses larger than $M$ (variance smaller than $S$) is then given by:
\ben
F(S) = 1 - \int_{-\infty}^{\deltac}\dd\delta \,p(\delta,S) ={\rm erfc}\( \frac{\Delta\deltac}{\sqrt{2\,\Delta S}}\)\,.
\een
This can be converted into the probability function to have a first up-crossing of the collapse barrier at variance (equivalently scale) $S$, given initial condition $\delta_0$ and $S_0$:
\ben
f(S|\delta_0,S_0) =\ddfrac{F}{S} = \dfrac{\Delta\deltac}{\sqrt{2\pi}\Delta S^{3/2}}
\exp\( -\dfrac{(\Delta\deltac)^2}{2\,\Delta S}\)  \,.
\een
The corresponding number (co)density (per comoving volume) of halos follows:
\ben
\label{eq:cosmoMF}
\ddfrac{n(M,z)}{M} = \dfrac{\rhomattO}{M}\,f(S|\delta_0,S_0) \,
\lb \ddfrac{S}{M}\rb\;.
\een
In the following, we will take $\delta_0=S_0=0$, setting initial conditions on very large scales. Probability function $f$ becomes:
\ben
f(S|\delta_0,S_0)\to f(S) = \dfrac{\deltac}{\sqrt{2\pi} S^{3/2}}
\exp\( -\dfrac{\deltac^2}{2\,S}\)  \,.
\een
The number density given in \citeeq{eq:cosmoMF} characterizes the cosmological halo mass function originally derived by Press~\&~Schechter (PS) \cite{PressEtAl1974}, but made fully consistent with the so-called cloud-in-cloud problem in the statistical EST approach. The Gaussian part of the probability function $f$ above is usually referred to as the PS function --- we will extend this terminology to the whole $f$ in the following. Other ansatze can be used, like the Sheth-Tormen (ST) function that was designed to better describe ellipsoidal collapse \cite{ShethEtAl1999,ShethEtAl2001}, or other variants \cite{ColeEtAl2008,ParkinsonEtAl2008}. Such slight modifications to the PS function will be used below.

The time or redshift dependence of the mass function comes from the growth of the contrast density field $\delta(z) = \delta(0)D(z)$ and of the variance $S(R)\propto D^2(z)$ as time goes. To explicitly account for it in the above development, it is sufficient to promote the fixed collapse condition \deltac\ to a redshift-dependent condition, while expressing the contrast variable $\delta$ and the variance $S$ associated with comoving scale $R$ (or mass $M$) at present time. This redshift-dependent threshold is defined as:
\ben
\delta_c(z)\equiv \dfrac{\deltac}{D(z)}  \;.
\een
This makes it clear why it is the present value of the matter density \rhomattO\ that appears in the mass function. In the following, each occurrence of \deltac\ should be understood as redshift dependent, unless specified otherwise.

The statistical description at the basis of the EST formalism can further be used to predict the mass function of subhalos in host halos. Indeed, it can be turned in terms of the conditional probability to have an up-crossing of the collapse barrier at variance $S_2$ given another up-crossing at variance $S_1$. The same formalism can be used to infer the number of halos of a given mass formed at some time, which are found contained in a bigger halo at a later time. Several approaches have been developed to approach the building of merger trees, \eg~\citerefs{Cole1991,LaceyEtAl1993,SomervilleEtAl1999,ColeEtAl2000,ColeEtAl2008,ParkinsonEtAl2008,JiangEtAl2014}. In the following, we follow the semi-analytical prescriptions presented in \citeref{HiroshimaEtAl2022}, and inherited from \citeref{YangEtAl2011} which inspired other works (\eg~\cite{JiangEtAl2020}). This very fast semi-analytical procedure allows us to derive a self-consistent subhalo mass function for a given host halo on a laptop at any redshift, given a set of input cosmological parameters. It was shown to reproduce results of cosmological simulations very well \cite{HiroshimaEtAl2022}.

In practice, we start the merger tree from the final stage of the host halo. For a host halo of mass $M(z)$ at redshift $z$ the number of subhalos of mass $m$ that accreted between $z$ and $z + \Delta z$ is given by: 
\ben
\label{eq:semiMF}
N(m) &=& \frac{M(z)}{m}\dfrac{\Delta_z\deltac}{\sqrt{2\pi}(\Delta_{mM}S)^{3/2}}\left|\frac{\mathrm{d}S}{\mathrm{d}m}\right|\\
&\times & G\left(\sqrt{\dfrac{S(M)}{S(m)}}, \frac{\delta_c(z + \Delta z)}{\sqrt{S(m)}}\right)\exp\left[-\frac{(\Delta_z\deltac)^2}{2(\Delta_{mM}S)}\right]\nn
\een
where $\Delta_z\deltac\equiv(\delta_c(z + \Delta z) - \delta_c(z))$ and $\Delta_{mM}S\equiv (S(m) - S(M))$. Function $G$ is defined as $G(x, y) = G_0x^{\gamma_1}y^{\gamma_2}$, with $(G_0, \gamma_1, \gamma_2) = (0.57, 038, -0.01)$, and was introduced in \citerefs{ParkinsonEtAl2008,ColeEtAl2008} to correct to nominal PS Gaussian function extracted from the EST formalism, and to get a good matching with cosmological simulation results. Out of these subhalos, the progenitor of the host is the most massive one. If the redshift step is short enough, there should always be a subhalo of mass $\tilde m>M(z)/2$, the host progenitor. To determine this mass, we compute the cumulative distribution function $C(\tilde m)$ of \citeeq{eq:semiMF} normalized in the range $[M/2, M]$:
\ben
C(\tilde m) = \frac{\int^{\tilde m}_{M/2}\dd m'\, N(m')}{\int_{M/2}^{M}\dd m'\, N(m')}\;.
\een
Drawing a random number $x\in[0, 1]$, we set the mass $\tilde m$ of the progenitor so that the cumulative distribution function up to this mass is equal to $x$, $C(\tilde m) = x$. The process is then repeated with $M(z+\Delta z)  = \tilde m$. Since this is a random process, we need to repeat it multiple times and take the median of the distribution of $M$. In practice, we run 500 realizations with 800 redshift steps between $z=10$ and $z=0$ (backward).

We can further construct the mass function counting in the same way the number of halos with masses between $m$ and $m + \dd m$ accreted onto the main progenitor between $z$ and $z+\Delta z$:
\ben
\ddfrac{N}{m} &=& \dfrac{\Delta z}{(1+z)} \dfrac{\Macc(z)}{m}\frac{1}{\sqrt{2\pi}}\frac{\Delta_z\deltac}{(\Delta_{mM}S)^{3/2}}\left|\frac{\mathrm{d}S}{\mathrm{d}m}\right|\\
&\times & G\left(\sqrt{\frac{S(M)}{S(m)}}, \frac{\delta_c(z + \Delta z)}{\sqrt{S(m)}}\right)\exp\left[-\frac{(\Delta_z\deltac)^2}{2(\Delta_{mM}S)}\right]\nn\;,
\een
with $\Macc(z) =( M(z) - M(z+\Delta z))$ the total mass accreted onto the main progenitor. We check the consistency of this mass function so that $\Macc(z) = \int\dd m\,m\,(\mathrm{d}N/\mathrm{d}m)$, and apply a correction when necessary (usually very small). Finally, we add up the contributions of all of the redshift bins to initial subhalo mass function of the main progenitor at the highest redshift of the run to get the full mass function at the final (lowest) redshift of interest.

\begin{figure}[!t]
\centering
\includegraphics[width=0.49\textwidth]{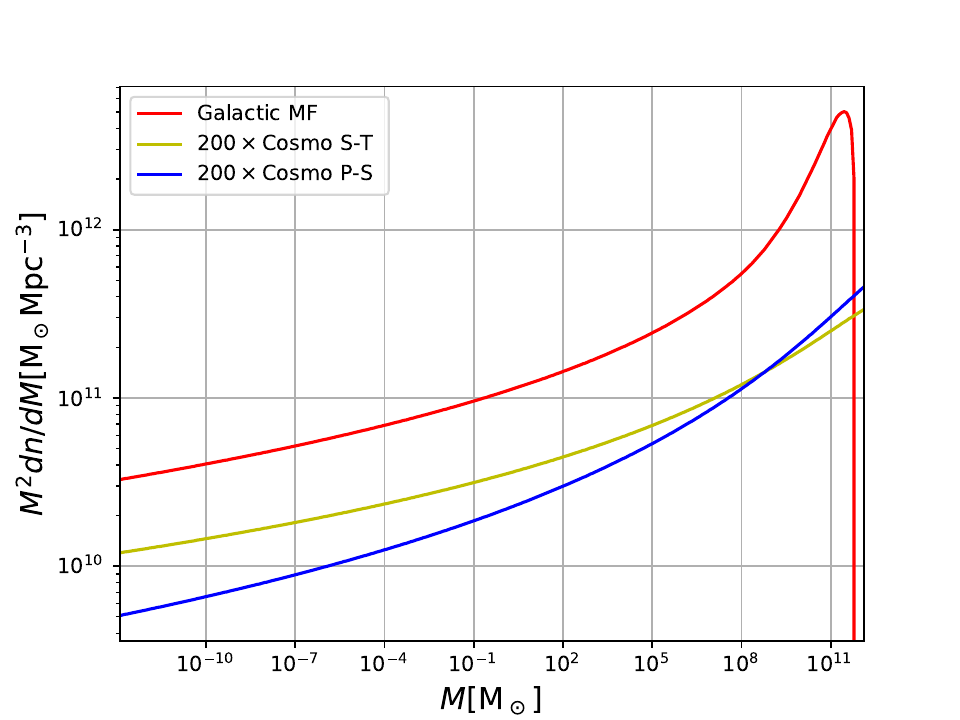}
\caption{\small Subhalo mass function predicted for a host halo similar to that of the Milky Way (red curve). It is compared with naive and simplistic estimates based on the cosmological halo mass function rescaled by a factor of 200, assuming either the PS spherical collapse (blue) or the ST ellipsoidal collapse (yellow).}
\label{fig:MWsubhaloMF}
\end{figure}

In \citefig{fig:MWsubhaloMF}, we show the resulting (averaged) subhalo mass function obtained for a host halo of mass $10^{12}\Msun$, a proxy for the Milky Way. It is compared with the cosmological mass function given in \citeeq{eq:cosmoMF} times a factor of 200, representative of a local overdensity at $z=0$ -- computed assuming both the PS \cite{PressEtAl1974} and ST \cite{ShethEtAl2001} distribution functions. Such a naive analogy or estimate is admittedly far from reliable, and the merger-tree approach is therefore key to soundly relate any subhalo population to a given cosmological setting.

\subsection{Adding an UCMH component}
So far, we have discussed the canonical DM case, with the variance $S$ stemming from a given matter power spectrum \mpspec. Such a description is perfectly fine to predict subhalo populations arising from the power-law part of the primordial power spectrum given in \citeeq{eq:pps}. However, there should be another population coming from the additional peak, reflecting the fact that the corresponding total matter power spectrum is now a sum of these two components:
\ben
\mpspec^{({\rm tot})}(k,z) = \mpspec^{(\text{p-l})}(k,z) + \mpspec^{(\dmosym)}(k,z)\;.
\een
Therefore, according to the definition of the variance of the density field $S$ given in \citeeq{eq:defS}, the presence of the additional peak also translates into a second component in the variance expression, such that:
\ben
S_{\rm tot}(R) = S_\text{p-l}(R) + S_\dmosym(R)\;.
\een
What is related to the power-law part $S_\text{p-l}(R)$ can be fully addressed by what has just been presented for the canonical dark matter subhalos. However, the peak part needs a specific treatment. The associated variance is readily calculated:
\ben
S_\dmosym(R) &=& \acleft \bar S_\dmosym \equiv \frac{4}{25} \left[\frac{D(0)k_s^2}{\OmattO H_0^2}\right]^2\mathcal{A}_\dmosym T_{\rm tr}^2(k_\dmosym)\left(\frac{k_\dmosym}{k_s}\right)^4\acright\nn\\
&&\times |\hat W_R(k_\dmosym)|^2\,,
\label{eq:Sdmo}
\een
where $k_s$ is the reference scale introduced in \citeeq{eq:pps} and only acts here as a normalization convenience, and where we recall that our window function $\hat W_R(k_\dmosym)$ is nothing but the Heaviside function.

\begin{figure*}[!t]
\centering
\includegraphics[width=0.49\textwidth]{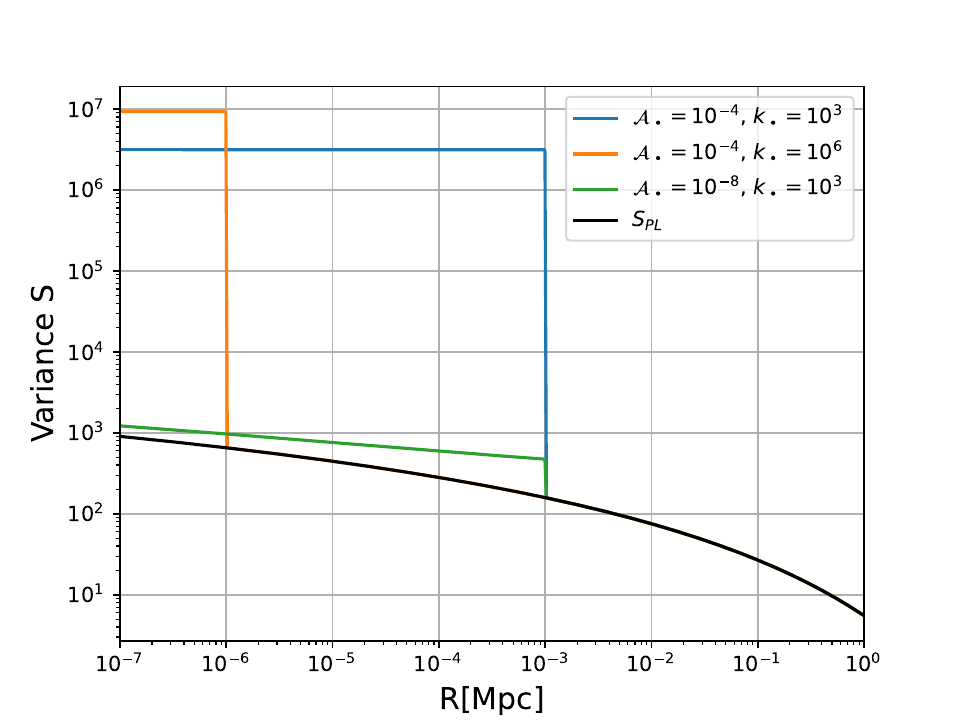}
\includegraphics[width=0.49\textwidth]{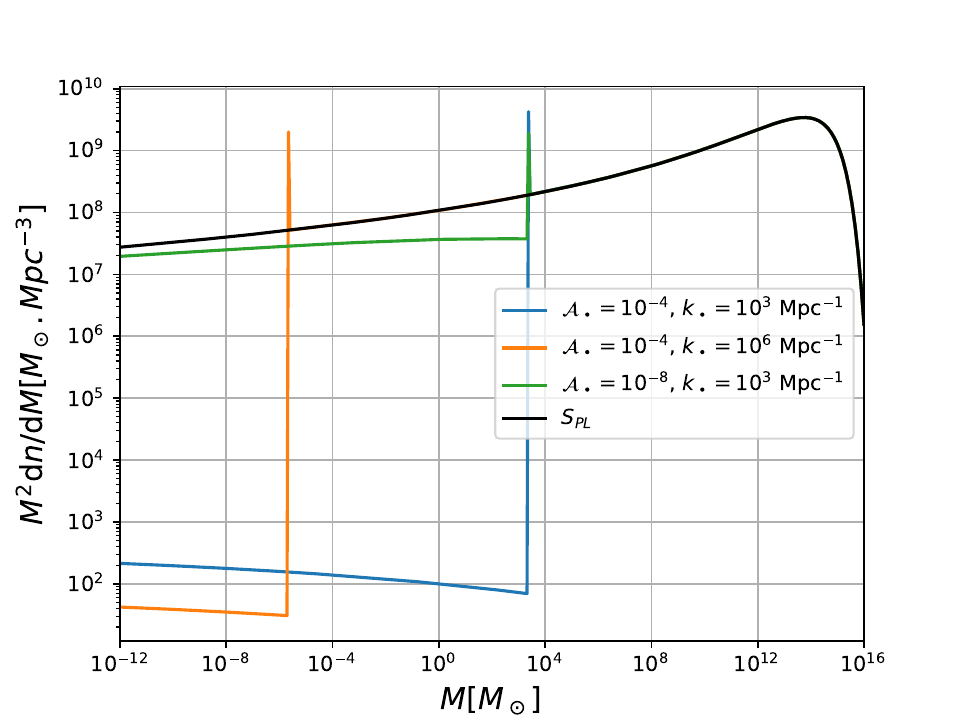}
\caption{\small {\bf Left panel:} Total variance of the matter power spectrum including the peak part of the primordial power spectrum, as a function of the spatial scale $R$ for different amplitudes $\mathcal{A}_\dmosym$ and injection scales $k_\dmosym$. Power on scales larger than $R_\dmosym = 1/k_\dmosym$ is unaffected by the peak while it is strongly enhanced on smaller scales. {\bf Right panel:} Corresponding cosmological mass functions, together with the power-law part of the power spectrum.}
\label{fig:variance_ucmh}
\end{figure*}

We report our calculation of $S_{\rm tot}$ as a function of scale $R=1/k$ in the left panel of \citefig{fig:variance_ucmh}, including the contribution of the peak component of the primordial power spectrum, where we see that it is strongly enhanced on scales smaller than $R_\dmosym=1/k_\dmosym$, which means that the collapse of structures can start very early on those scales --- these are the UCMHs. This also means that all structures smaller than $R_\dmosym$ cross the collapse barrier at the same time, which implies that only the bigger ones, directly fixed by $R_\dmosym$, matter (they contain the smaller ones). However, the determination of the associated mass function is more complicated because of the derivative of $S_\dmosym$ which is not well defined at $k_\dmosym$, as a consequence of $\hat W$ being a Heaviside function. It was actually shown through a detailed analysis in \citeref{AbellanEtAl2023} that in this case, the cosmological mass function given for the canonical case in \citeeq{eq:cosmoMF} can be reformulated as:
%
%\ddfrac{n(M,z)}{M} = \dfrac{\rhomattO}{M}\,f(S|\delta_0,S_0) \,
%\lb \ddfrac{S}{M}\rb\;.
%
\ben
\ddfrac{n(M, z)}{M} = \dfrac{\rhomattO}{M}
\acleft f(S) \,\lb \ddfrac{S_\text{p-l}}{M} \rb  + \delta(M-M_\dmosym)\,\mathcal{F}_\dmosym(z)
\acright ,\nn\\
\label{eq:dndm_dmo}
\een
where it is clear that the collapsed mass of UCMHs is uniquely determined by the injection scale $k_\dmosym$, and where their fraction is given by:
\ben
\mathcal{F}_\dmosym(z) = {\rm erf}\left(\frac{\nu_+(z)}{\sqrt{2}}\right) -
        {\rm erf}\left(\frac{\nu_-(z)}{\sqrt{2}}\right)\;,
\een
with        
\ben
\nu_+(z) &\equiv & \frac{\deltac(z)}{\sqrt{S_\text{p-l}(M_\dmosym)}}\\
\nu_-(z) &\equiv & \frac{\deltac(z)}{\sqrt{S_\text{p-l}(M_\dmosym) + \bar S_\dmosym}}\nn\;,
\een
where $\bar S_\dmosym$ has been implicitly defined in \citeeq{eq:Sdmo}. The derived cosmological mass function is shown in the right panel of \citefig{fig:variance_ucmh}.

We emphasize that since the peak part of the variance is way higher than its power-law counterpart, the associated UCMHs form much earlier, hence in a much denser universe. This also impacts their internal properties, as will be discussed below.

To go from the cosmological mass function to the subhalo mass function in a host halo, we can essentially follow the same procedure as developed above, with minor adjustments --- all this goes beyond the work of \citeref{AbellanEtAl2023}, which is limited to the cosmological mass function. To compute the mass fraction of UCMHs in the host halo, we have to integrate all the contributions of those UCMHs accreted during the merging history:
\ben
\mathcal{F}_{\dmosym}^{\rm host} = \frac{1}{M}\int_{0}^{z_{\rm tree}} \dd z \,\mathcal{F}_\dmosym (z)M_{\rm acc}(z)   
\een
where $M_{\rm acc}(z)$ is the mass accreted by the host halo between $z$ and $z+\dd z$, and where the integral (or weighted sum) is performed between the redshift of interest, here $z=0$, and the highest redshift $z_{\rm tree}$ fixed by the balance between a good mass resolution for seed of the merger tree and a decent calculation time (as already mentioned above, we take $z_{\rm tree}=10$).

\begin{figure*}[!t]
\centering
\includegraphics[width=0.49\textwidth]{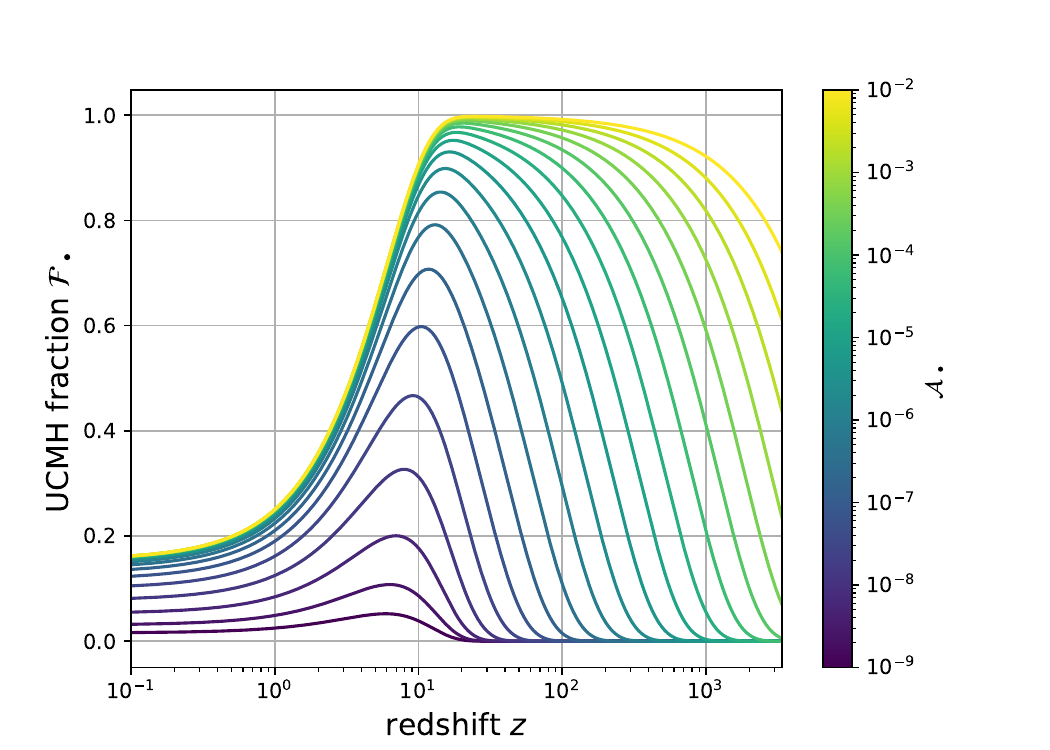}
\includegraphics[width=0.49\textwidth]{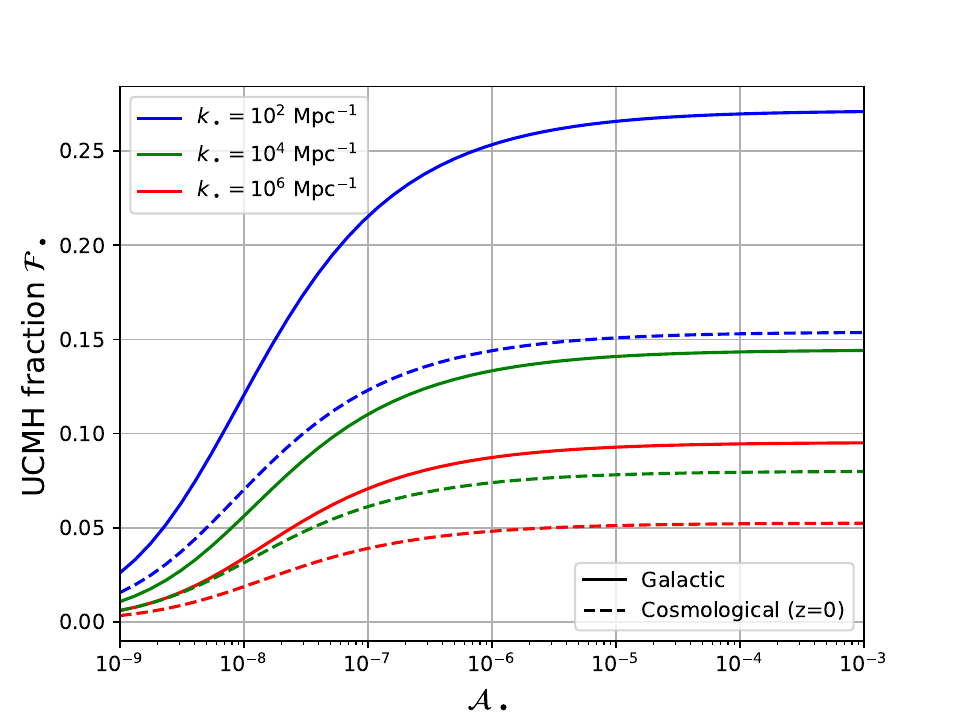}
\caption{\small {\bf Left panel:} Cosmological fraction of UCMHs as a function of redshift, for $k_\dmosym = 10^2 {\rm Mpc}^{-1}$ $(M_\dmosym = 2.3 \times 10^3 \Msun)$, and for different peak amplitudes $\mathcal{A}_\dmosym$ (color code). {\bf Right panel:} Comparison between the UCMHs fraction in a Milky Way-like host halo and the cosmological fraction, as a function of the peak amplitude $\mathcal{A}_\dmosym$  and for different injection scales $k_\dmosym$.}
\label{fig:fucmh}
\end{figure*}

In \citefig{fig:fucmh}, we show in the left panel the cosmological fraction of UCMHs as a function of redshift for different peak amplitudes, assuming an injection scale of $k_\dmosym = 10^3 {\rm Mpc}^{-1}$ $(M_\dmosym = 2.3 \times 10^3 \Msun)$ --- recovering here results obtained in \citeref{AbellanEtAl2023}. We further compare in the right panel the UCMH fraction in a Milky Way-like host halo with the cosmological one as a function of peak amplitude, for different injection scales. For the total cosmological fraction, we simply take $\mathcal{F}_\dmosym(z=0)$.

Increasing $\mathcal{A}_\dmosym$ increases the number of UCMHs up to the point where all DM is made of UCMHs at their formation redshift, and then it is constant. As scales larger than $M_\dmosym$ collapse, some of the UCMHs will merge to form standard halos before further merging with the host halo, and we conservatively discard them, considering they are part of bigger subhalos. This conservative picture implies that small-scale UCMHs are less likely to be accreted by a Milky Way-like host halo than bigger ones, which translates into a decrease of $\mathcal{F}_{\dmosym}^{\rm host}$ with increasing $k_\dmosym$, as visible in the left panel of \citefig{fig:fucmh}. We remind that our final subhalo mass function, including UCMHs, was presented earlier in the core of the paper, in \citefig{fig:UCMH_mf}, assuming a Milky Way-like host halo of $10^{12}\,\Msun$. This figure was actually made from all of the ingredients introduced just above in this appendix section.

\subsection{Basic tidal evolution of subhalos and UCMHs}
Since we have defined the subhalo mass function for both standard subhalos and UCMHs in a given host halo, we further need to specify how these objects are spatially distributed and evolve in their host halo, assumed spherically symmetric. Here, we will follow the semi-analytical approach developed in \citerefs{StrefEtAl2017,HuettenEtAl2019,FacchinettiEtAl2022}. We suppose that all subhalos first behave like hard spheres and spatially track the overall DM distribution at accretion, before experiencing tidal stripping. For the Milky Way DM halo, we assume the best-fitting NFW model of \citeref{McMillan2017}, which is constrained on kinematic data, and for which the local density of MD at the solar position is of order 0.01~$\Msun/{\rm pc}^3$. Due to tidal stripping, the initial masses subhalos have at accretion are modified as a function of their orbits --- for simplicity, we assume circular orbits, merely characterized by the radial distance to the host halo center. We further apply a criterion for tidal disruption explained below. All this implies that the initial mass function is modified and becomes radial dependent. We first introduce the generic way we implement tidal stripping in our calculation, before giving details about the internal properties of subhalos and UCMHs which will determine their response to tidal stripping.

\begin{figure*}[!t]
\centering
\includegraphics[width=0.49\textwidth]{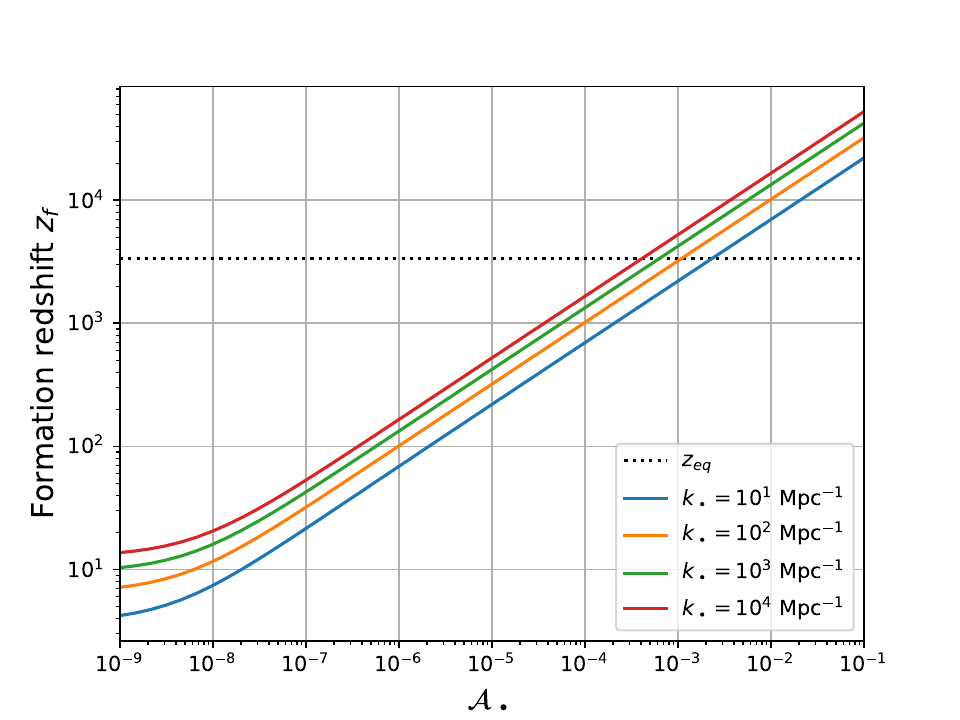}
\includegraphics[width=0.49\textwidth]{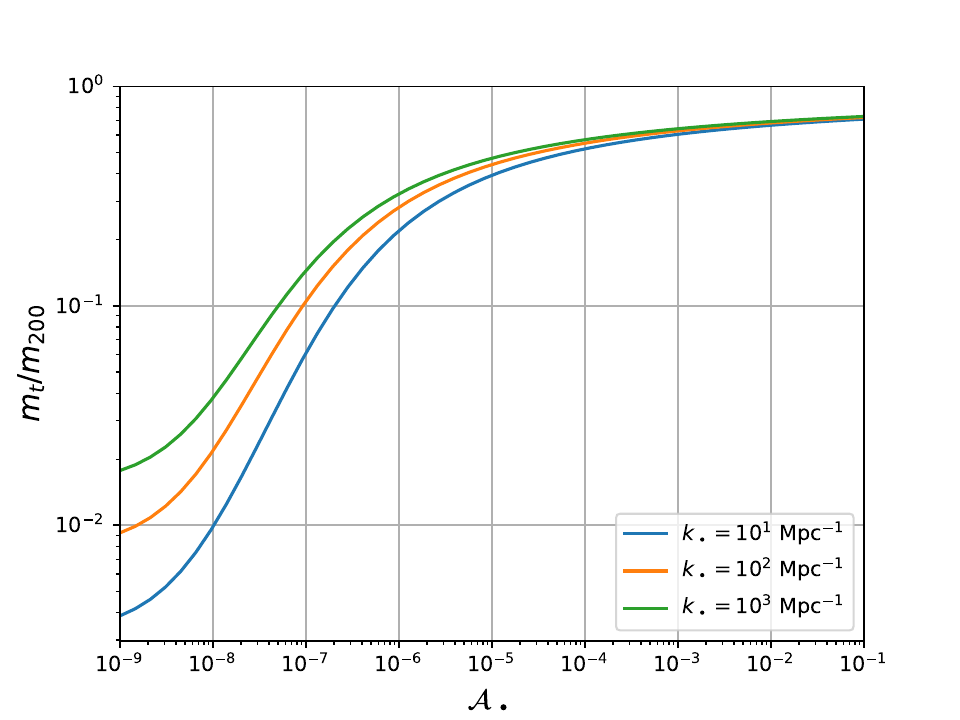}
\caption{\small {\bf Left panel:} UCMH formation redshift as a function of the spike's amplitude $\mathcal{A}_\star$. {\bf Right panel:} Ratio of tidal-to-initial mass as a function of the peak amplitude for UCMHs.}
\label{fig:zfUCMH}
\end{figure*}

In this work, we simply consider the tidal stripping induced by the gravitational potential of the host halo (baryonic effects are more involved, but expected to lead to minor changes for subhalos passing by the solar neighborhood --- see \citeref{FacchinettiEtAl2022}). The tidal radius of a subhalo (or UCMH), \rtide, is defined as the maximal distance from the center of the subhalo for which a test particle remains bound and does not leak away to the host halo. It can be defined as the Roche radius given by \cite{BinneyEtAl2008}:
\ben
\rtide &=& R \left[\frac{m(\rtide)}{3M(R)f(R)} \right]^{1/3} \\
\text{with} \quad f(R) &=& 1 - \dfrac{1}{3}\ddfrac{\ln M}{\ln R}\,.\nn
\een
Here $R$ is the distance from the center of the host, not to be confused with the comoving scale used when reviewing structure formation above, and $M(R)$ is the enclosed mass of the host halo. The subhalo enclosed mass $m(r)$ is derived from its mass density profile, which we can describe from the generic power-law density function \cite{Zhao1996}:
\ben 
\label{eq:profile}
\rho(x) = \rhoscale \left\{g(x) \equiv \frac{1}{x^\gamma(1+x^\alpha)^{(\beta-\gamma)/\alpha}}\right\}\,,
\een
where $x = r/\rscale$, $\rscale$ and $\rhoscale$ are the scale radius and density, and $g$ is the dimensionless mass density. Index $\gamma$ sets the inner slope ($\gamma =0$ leads to a cored halo), $\beta$ the outer slope and $\alpha$ controls how fast the profile transitions between the two asymptotic regimes. The NFW profile \cite{NavarroEtAl1996a}, which we will use for standard DM subhalos, is recovered with $(\alpha, \beta, \gamma) = (1, 3, 1)$. The enclosed mass follows:
\beq
m(x) = 4\pi \rhoscale \rscale^3\left\{\mu\left(x\right) \equiv \int_0^{x}\dd x'\,x'^2\,g(x')\right \}
\eeq
where $\mu$ is the dimensionless enclosed mass. The virial extent $\rvir$ of a structure assumed isolated in a flat background is usually defined so that its average density is 200 times the critical density, which also relates to the concept of concentration $\cvir\equiv \rvir/\rscale$. Therefore, a couple $(\rhoscale,\rscale)\leftrightarrow(\mvir=m(c);\cvir)$ fully determines the inner properties of a halo. The dimensionless tidal radius $\xtide\equiv \rtide/\rscale$ can therefore be expressed as a function of the scale density $\rhoscale$ :
\ben
 \frac{\mu(\xtide)}{\xtide^3} = \frac{3M(R)f(R)}{4\pi \rhoscale R^3}\,,
\een
and some mass is tidally lost to the host halo as soon as $\xtide<\cvir$.

The tidal mass $\mtide = m(\rtide)$ is the mass left after tidal stripping, and therefore the one that we consider for heating. We also implement a tidal disruption criterion set by the condition $\xtide<\mymax{x}$, with the critical value $\mymax{x}$ possibly varying between virtually $\ll 1$ for resilient halos \cite{vandenBoschEtAl2018,vandenBoschEtAl2018a,ErraniEtAl2020} up to $\sim 1$ for fragile halos \cite{TormenEtAl1998,HayashiEtAl2003,DiemandEtAl2004,DiemandEtAl2008b,SpringelEtAl2008,vandenBosch2017} --- see \citerefs{StrefEtAl2017,FacchinettiEtAl2022} for details. Note that for UCMHs, there is essentially no tidal disruption given their compactness.

In this paper, we are mostly interested in UCMHs, whose scale density \rhoscale\ is expected to be much larger than that of standard subhalos, because they form much earlier. This higher ``compactness'' is key to maximize the heating rate, and we actually found that standard subhalos have only minor effects on the stability of exoplanetary systems with respect to UCMHs.

Simulations indicate that the mass density in UCMHs follows a Moore profile \cite{MooreEtAl1998}, corresponding to $(\alpha, \beta, \gamma) = (1, 3, 3/2)$ in \citeeq{eq:profile}. The scale parameters obtained in dedicated simulations are given by \cite{DelosEtAl2018,DelosEtAl2018a}:
\ben
\label{eq:ucmh_props}
\rscale &=& f_1 k_\dmosym^{-1}(1+z_{\rm f})^{-1}\\
\rhoscale &=& f_2 \rhomattO(1+z_{\rm f})^3\nn\,,
\een
with $f_1 = 0.7$ and $f_2 = 30$, and $z_{\rm f}$ the redshift of formation of the UCMH. For the counting in the mass function, we match the mass $m_{200}$ of the UCMH with the mass $m$ set by $k_\dmosym$ in the EPS approach introduced above, hence the departure of iso-mass curves from vertical lines in \citefig{fig:limits}.

The formation redshift $z_{\rm f}$ is defined as the redshift at which the variance of the matter power spectrum equals to the collapse barrier, $S(m) = \deltac^2(z_{\rm f})$. At that redshift, the corresponding region of space is statistically more likely to collapse. This turns out to be a very good approximation to the actual sharp distribution of formation redshifts \cite{DelosEtAl2018,DelosEtAl2018a}. A larger amplitude of the peak $\mathcal{A}_\dmosym$ in the primordial power spectrum leads to a larger variance and thus a larger formation redshift, as illustrated in in the left panel of \citefig{fig:zfUCMH}.

The earlier the formation of UCMHs, the more compact and dense they are, and therefore the more resilient to tidal stripping. This is shown in the right panel of \citefig{fig:zfUCMH} through the ratio of tidal-to-initial mass.

This compactness strongly increases the heating rate compared to NFW halos because (i) the UCMH tidal masses are much higher than classical NFW subhalos at a given  galactocentric radius $R$, and (ii), being smaller in size for a given mass, closer encounters are more likely, leading to more efficient heating.

All of the developments above are necessary to go from the heating constraints up to the fundamental cosmological parameters $\mathcal{A}_\dmosym$ and $k_\dmosym$ of the UCMH model.

\section{From adiabatic invariant actions to the heating-induced evolution of planet orbital parameters}
\label{app:orbits}
Here, we relate the time evolution of the planet's adiabatic invariants (actions) to the impulsive heating from DMO encounters. This allows us to connect the evolution of the orbital parameters to the heating.

We first define the radial action for the planet as:
\ben
J_r &\equiv& \dfrac{1}{2\pi}\oint\dd r \, p_r= \dfrac{1}{\pi}\int_{r_-}^{r_+}\dd r\,p_r \\
 &=& \dfrac{1}{\pi}\int_{r_-}^{r_+}\dd r\,\sqrt{2\left(E - \frac{L^2}{2r^2} + \frac{G\mstar}{r}\right)}\,,\nn
\een
where $E$ is the total energy and $L$ the angular momentum of the planet in the star frame (we assume $\mplan\ll\mstar$, but a reformulation in the center-of-mass frame would be straightforward). By definition, the apocenter and pericenter, $r_-$ and $r_+$,  are the roots of the integrand:
\ben
r_\pm = \dfrac{G\mstar \pm\sqrt{G^2\mstar^2 - 2 \mE L^2}}{2\mE}\,,
\een
where $\mE\equiv-E>0$, which translates into:
\ben
J_r &=&\dfrac{1}{\pi} \int_{r_-}^{r_+}\dfrac{\dd r}{r}\sqrt{2\mE (r_+ - r)(r - r_-)}\\
&=& \frac{1}{2}\sqrt{2\mE}\left(r_+ - r_- - 2\sqrt{r_+r_-}\right)\nn\\
&=&\frac{G\mstar}{\sqrt{2\mE}} - L\nn\,.
\een
We can link this expression to the eccentricity $e$ through the relations:
\ben
\label{eq:ELexc}
\mE = \frac{G\mstar}{2\,a}\quad, \quad L = \sqrt{G\mstar a(1-e^2)}\,,
\een
valid for a Keplerian potential, where $a$ is the semi-major axis. We get:
\ben
J_r &=& \sqrt{G\mstar a}\left(1 - \sqrt{1-e^2}\right)\\
&\underset{e\to 0}{\sim}&\frac{1}{2} e^2\sqrt{G\mstar a}\,.\nn
\een
The time variation of the action is related to the heating rate through:
\ben
%\ddfrac{J_r}{t} = \dfrac{P}{4\pi}\Heatplan = \dfrac{1}{2}\sqrt{\frac{a^3}{G\mstar}}\Heatplan\,.
\ddfrac{J_r}{t} = \dfrac{P}{2\pi}\Heatplan = \sqrt{\frac{a^3}{G\mstar}}\Heatplan\,.
\een
This leads to:
\ben
\label{eq:evol_a_e}
\dfrac{\dot{(e^2)}}{\sqrt{1-e^2}} + \dfrac{\dot a}{a}\left( 1-\sqrt{1-e^2}\right) = \dfrac{2\,a\,\Heatplan}{G\mstar}\,.
\een
If we consider a planet on a constant semi-major axis, $\dd a/\dd t = 0$, such that $a\sim\Rplan$, and very low eccentricity, $e\ll 1$, then the eccentricity relates directly to the heating rate as:
\ben
\label{eq:e_approx}
e \sim \sqrt{\frac{2\,\mathcal{T}\,\Rplan\,\Heatplan}{G\mstar}} = \sqrt{\frac{E_{\rm inj}}{E_{\rm bind}}}=\sqrt{\Rheat}\,,
\een
where the heating ratio \Rheat\ was already defined in \citeeq{eq:Rheat}. This very crude estimate suggests that an increase in the eccentricity can be related to heating processes, a common feature of radial migration \cite{AumerEtAl2016}.

To solve this problem more rigorously, we have to close \citeeq{eq:evol_a_e}. We can do so by working out the change in the angular momentum due to a velocity kick $\dvvplan$: 
\ben
\myvec{\delta L} = \myvec{r}\times \dvvplan\,,
\een
where $\myvec{r}$ is the position of the planet in the star frame. The resulting change $\Delta L$ in the norm of $\myvec{L}$ follows:
\ben
\Delta L &=& |\myvec{L} + \myvec{\delta L}| - |\myvec{L}|\\
 &=&\frac{\myvec{L}\cdot\myvec{\delta L}}{L} + \dfrac{1}{2}\left[\dfrac{(\myvec{\delta L})^2}{L} - \dfrac{(\myvec{L}\cdot\myvec{\delta L})^2}{L^3}\right] + \mathcal{O} \left(\frac{\delta L^3}{L^3}\right)\,.\nn
\een
Averaging over many encounters makes the first-order term vanish, assuming $\dvvplan$ is isotropically distributed. For the second-order terms, we get:
\ben
\myav{(\myvec{\delta L})^2} &=& \myav{\myvec{r}^2\dvvplan^2 - (\myvec{r}\cdot\dvvplan)^2}
= \frac{2}{3}r^2\,\dvplan^2\,,
\een
where the average is computed over all directions in a sphere, with $\myav{\cos^2\theta} = 1/3$. The other term becomes:
\ben
\myav{\myvec{L}\cdot\myvec{\delta L})^2} & = & \myav{(\dvvplan\cdot(\myvec{L} \times \myvec{r}))^2}
= r^2L^2\myav{(\dvvplan\cdot\myvec{e}_\phi)^2}\nn\\
&=&\frac{1}{3}L^2r^2\dvplan^2\,,
\een
using the same averaging procedure. Unit vector $\myvec{e}_\phi$ characterizes the direction of the unperturbed planet velocity. Combining both results we obtain:
\ben
\Delta L = \dfrac{r^2\dvplan^2}{6L} = \dfrac{r^2}{3L}\Heatplan\,\delta t\,.
\een
To get the value of $\Delta L$ averaged over an orbit, we can write $r$ as a function of the eccentric anomaly $u$:
\ben
r = a(1-e\cos(u))\,,
\een
and further make a change of variable between $u$ and time $t$ using:
\ben
t = \dfrac{P}{2\pi}(u - e\sin u)\,,
\een
where $P$ is the orbital period. Consequently:
\ben
\dd t = \dfrac{P}{2\pi}\dfrac{r}{a}\dd u\,.
\een
The orbital average of $r^2$ is then given by:
\ben
\myav{r^2}_\oslash &=& \dfrac{1}{P} \int_0^P\dd t\,r^2(t)
=\dfrac{1}{2\pi}\int_0^{2\pi}\dd u\,\dfrac{r^3(u)}{a}\\
&=&\dfrac{a^2}{2\pi}\int_0^{2\pi}\dd u\, (1 - 3e\cos u + 3e^2 \cos^2 u - e^3\cos^3 u)\nn \\
& = & a^2 \left( 1+\dfrac{3}{2}e^2\right)\,.\nn
\een
Thus, we finally get:
\ben
\ddfrac{L}{t} = \dfrac{a^2\Heatplan}{3L} \left( 1+\dfrac{3}{2}e^2\right)\,.
\een

Therefore, we can now close \citeeq{eq:ELexc} with another complementary differential equation relating $a$ and $e^2$: 
\ben
\label{eq:res_dL}
\dfrac{\dot a }{a}(1-e^2) - \dot{(e^2)} = \dfrac{2\,a\,\Heatplan}{3G\mstar}\left(1+\dfrac{3}{2}e^2 \right)\,.
\een
Thus, contrary to what was naively expected from \citeeq{eq:e_approx}, for $e\ll1$ and a constant semi-major axis $a = \Rplan$, we get a negative $\dot{(e^2)}$:
\ben
\dot{(e^2)} \sim -\dfrac{2\,\Rplan\,\Heatplan}{3G\mstar}\,.
\een
This actually comes from the fact that the variation of $a$ is the main driver of the variation of $L$, so we cannot assume $a$ to be constant while varying $e$.

Combining \citeeq{eq:evol_a_e} and \citeeq{eq:res_dL} leads to the following system of equations, which is one of the main results of this part:
\ben
\label{eq:evol_a_e_full}
\begin{cases}
\dfrac{\dot a}{a} &= \dfrac{2\,a\,\Heatplan}{G\mstar} \left(1 + \dfrac{2+3\,e^2}{6\sqrt{1-e^2}}\right)\\
\dot{(e^2)} &= \dfrac{2\, a\, \Heatplan}{G\mstar} \left[\dfrac{2}{3} - \dfrac{3e^2}{2} + \left(\dfrac{1}{3} + \dfrac{e^2}{2} \right)\sqrt{1-e^2}\right]
\end{cases}\,.
\een
% HERE JL
If the eccentricity is large enough ($e\geq 0.8337$), $\dot{(e^2)}$ can be negative, which means that the heating tends to circularize the motion by increasing the angular momentum. In the limit of low eccentricity, this system reduces to:
\ben
\label{eq:evol_a_e_limit}
\begin{cases}
\dfrac{\dot a}{a} &= \dfrac{8\,a\,\Heatplan}{3G\mstar}\\
\dot{(e^2)}  &= \dfrac{2\,a\,\Heatplan}{G\mstar}
\end{cases}\;.
\een
There is an extra factor of $4/3$ compared to a quick estimate of the relation between the rate of change of radius of a circular orbit and the heating rate.\footnote{Indeed, a simple estimate can relate the time evolution of the energy of the circular orbit, $\mE = G\mstar/2\Rplan$, to the heating rate as follows: $\dd\mE/\dd t = \dd \Heatplan/\dd t$. This readily translates into a time evolution of the circular orbit which reads: $\dd\Rplan/\dd t = 2\Rplan^2\Heatplan/G\mstar$.} For a constant heating rate, these equations have an analytical solution:
\ben
%\begin{cases}
a(t) = \dfrac{a_0}{1 - t/\tej}\;,\;
\text{with}\;\tej \equiv \dfrac{3G\mstar}{8a_0\Heatplan}\,,
%\end{cases}\;,
\een
where $a_0$ is the initial semi-major axis. This solution is defined for $t\leq\tej$, where $\tej$ is the time at which the semi-major axis diverges and the planet is ejected. We can recast this result in terms of a time-dependent heating-to-binding energy ratio \tRheat\ as:
\ben
\label{eq:a_ana}
a(t) &=& \dfrac{a_0}{1 - \dfrac{4}{3}\tRheat(t)}\\
\text{with}\;\tRheat(t) &\equiv& \dfrac{2\,a_0\Heatplan\,t}{G\mstar}\nn\,.
\een
The planet is therefore ejected after the total injected energy has become larger than the binding energy, as we could have naively expected (note however the additional factor of $4/3$ to the naive reasoning, which means that heating by only 3/4 of the binding energy is enough to eject the planet).

Consequently:
\ben
\label{eq:e_ana}
e^2(t) = e_0^2 - \dfrac{3}{4}\ln\left(1-\frac{t}{\tej}\right) = e_0^2 + \dfrac{3}{4} \ln \left( \dfrac{a(t)}{a_0} \right)\,.
\een
This implies that the eccentricity increases rapidly. If we double the size of the semi-major axis, the eccentricity increases up to $\sqrt{(3/4)\ln(2)}\approx 0.72$.

To first order in $e^2$, the system reads
\ben
\begin{cases}
\dfrac{\dot a}{a} &= \dfrac{8\,a\,\Heatplan}{3G\mstar}
\left( 1+\dfrac{e^2}{2} \right) \\
\dot {(e^2)} &= \dfrac{2\,a\,\Heatplan}{G\mstar} \left(1 - \dfrac{7}{6} e^2 \right)
\end{cases}\;,
\een

We can further apply the same treatment to the vertical action, now in a spherical coordinate system rather than an axisymmetric one: 
\ben
J_z =J_\theta \equiv \dfrac{1}{2\pi}\oint \dd\theta\, p_\theta \,,
\een
where $\theta$ is the angle between the normal to the initial orbital plane and the position of the planet. In terms of the angles in spherical coordinates, the total angular momentum reads:
% HERE JL
\ben
L^2 = (r^2\dot\theta )^2 + (r^2\dot\phi)^2 = p_\theta^2 + \frac{L_z^2}{\sin^2\theta}\;.
\een
Conservation of $L$ gives:
\ben
p_\theta = \sqrt{L^2 - \frac{L_z^2}{\sin^2\theta}}\,.
\een
The motion is bounded by the points at which $p_\theta = 0$, which further translates into $\sin\,\theta_\pm = L_z/L = \cos \zeta_i$, where $\zeta_i$ is the inclination of the planet's orbit defined as the maximal angle between the initial orbital plane and the position of the planet. The vertical action becomes:
\ben
J_z = \dfrac{1}{\pi}\int_{\pi/2 - \zeta_i}^{\pi/2 + \zeta_i}\dd\theta\,\sqrt{L^2 - \dfrac{L_z^2}{\sin^2\theta}}\,.
\een
With a change of variable $\psi\equiv \pi/2 - \theta$, we finally get:
\ben
J_z &=& \frac{L}{\pi}\int_{-\zeta_i}^{\zeta_i}\dd\psi\,\sqrt{1-\dfrac{\cos^2\zeta_i}{\cos^2\psi}}
= L(1-|\cos \zeta_i|)\nn\\
&=& L - |L_z|\,.
\een
We can again relate the change in the vertical action to the impulsive (non-adiabatic) heating rate:
\ben
\ddfrac{J_z}{t} = \dfrac{P}{2\pi}\Heatplan\,,
\een
and then compute the time derivative of $J_z$ from the expression just above: 
\ben
\begin{cases}
\dot J_z &= \dot L(1-|\cos \zeta_i|) + L\dot \zeta_i s_i \sin \zeta_i\\
\dot L &=\dfrac{L}{2}\left(\dfrac{\dot a}{a} - \dfrac{\dot{(e^2)}}{1-e^2}\right)
\end{cases}\;,
\een
where $s_i\equiv{\rm sign}(\cos\zeta_i)$. We can safely consider only positive values of $\cos\zeta_i$ in the following.

Further using the previous results of radial heating for the evolution of $\dot a/a$ and $\dot{(e^2)}$ given in \citeeq{eq:evol_a_e_full}, we obtain our last main result on the evolution on the inclination angle:
\ben
\label{eq:evol_zeta_full}
\dot \zeta_i\sin \zeta_i = \dfrac{a\,\Heatplan}{G\mstar}\left[\dfrac{1}{\sqrt{1-e^2}} - \dfrac{2+3e^2}{6(1-e^2)}(1-\cos \zeta_i)\right].
\een
This differential equation can be solved numerically as an additional item to the system of equations in \citeeq{eq:evol_a_e_full}. To still get analytical insight, we can take the limit of low eccentricity and low inclination, which leads to:
\ben
\label{eq:zeta_ana}
\dot{(\zeta_i^2)} = \frac{2\,a\,\Heatplan}{G\mstar}\,.
\een
This is similar to the heating effect on $e^2$ obtained in \citeeq{eq:evol_a_e_limit}. The solution reads:
\ben
\label{eq:zeta_exact}
\zeta_i^2(t) = \zeta_{i,0}^2 + \dfrac{3}{4}\ln\left(\dfrac{a(t)}{a_0}\right)\,.
\een

\section{A closer look to our exoplanet data sample}
\label{app:data}

\begin{figure*}[!t]
\centering
\includegraphics[width=0.49\textwidth]{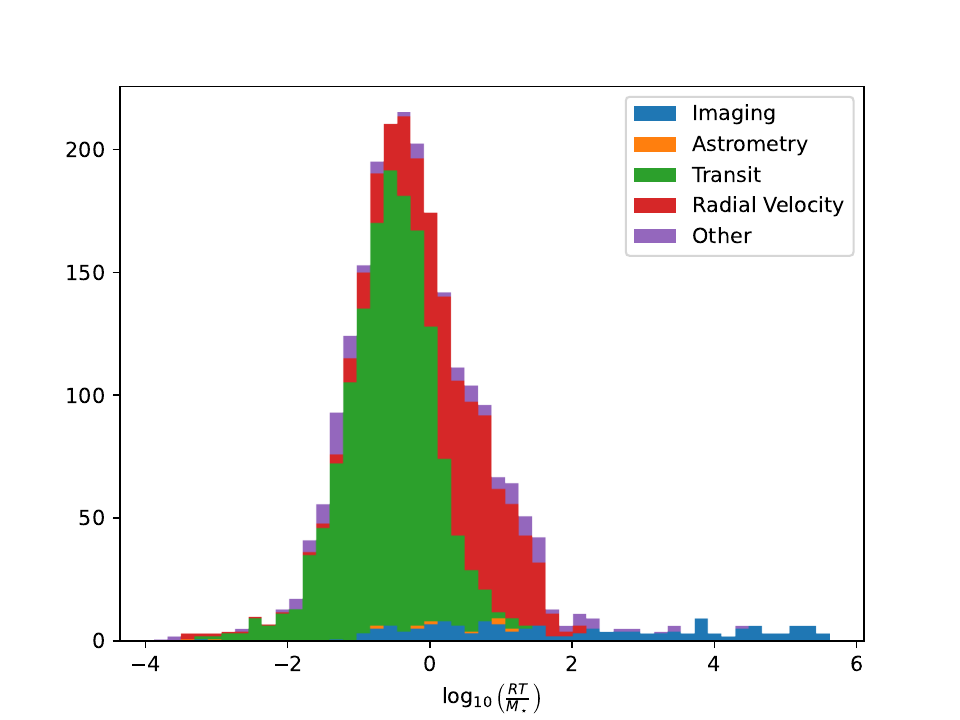}
\includegraphics[width=0.49\textwidth]{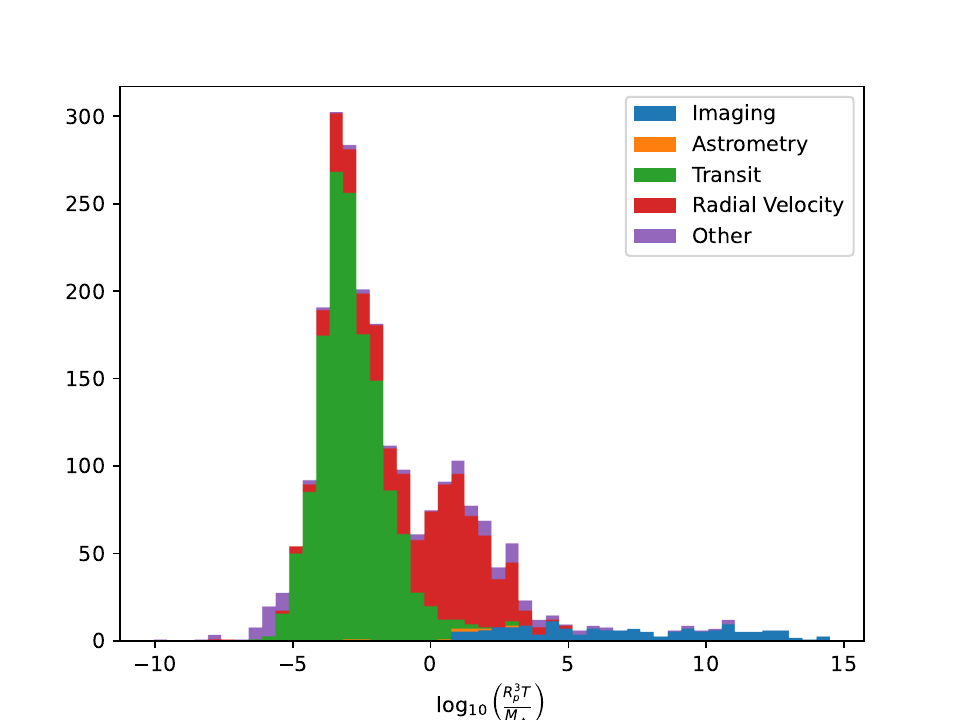}
\caption{\small {\bf Left panel}: Distribution of detected exoplanets as a function of the selection variable tuned for light DMOs. Different colors correspond to different detection methods as indicated by the legend. {\bf Right panel}: Same for the criterion tuned for heavier DMOs.}
\label{fig:histplanets}
\end{figure*}

This appendix section provides additional details about the exoplanet sample used in our analysis. We refer to \citeref{DeegEtAl2026}, and in particular to \citeref{Mordasini2018} (on top of additional references within), for a more exhaustive view on exoplanetery science.

In \citefig{fig:histplanets}, we show the distributions of detected exoplanets according to the two possible selection parameters we identified in \citesec{sec:constraints} -- in the left panel for a criterion tuned for light DMOs, and in the right panel for a criterion tuned for heavier ones. We also make explicit the detection methods associated with the available sample of planets, which reveals a strong segregation as a function of selection criteria. The most efficient way to detect exoplanets is the transit method, but this requires that the star, the planet, and the observer are aligned well enough to observe a dip in the star's luminosity. This method also requires observing multiple orbital periods. Obviously, this kind of observational limitation biases the detection toward planets close to their star and thus with low constraining power on external impulsive heating. On the other hand, direct imaging of exoplanets is way less efficient than the transit method but can instead capture planets very far from their star, which actually requires a minimal angular separation \cite{DeaconEtAl2014,ZhangEtAl2021,RothermichEtAl2024,CifuentesEtAl2025,Zurlo2026,Chauvin2023}. As we see, this population of distant planets globally manifests itself as a long tail in the overall distribution, which is the one relevant to our analysis.

To set constraints, we implicitly assume that the observed population of exoplanet is somewhat representative of the actual population. Actually, there is no predictive model for the distribution of exoplanet properties \cite{Mordasini2018}. One could therefore object that the observed distribution of planets could already be the result of the dynamical heating induced by a population of DMOs, which would eject most distant planets, which might affect our statistical strategy to set limits. A first way to counter-object is to remark that the distribution of directly imaged exoplanets is rather flat in both of these representations. We can still try to go beyond with a more quantitative discussion.

To better understand the effect of heating on the exoplanet population, we generated a population of exoplanets where the initial star mass is chosen following Salpeter's law \cite{Salpeter1955}:
\ben
\ddfrac{N}{\mstar}\propto \mstar^{-2.35}\,.
\een
We consider a flat distribution of age, corresponding to a constant star-formation rate. We then discard stars that would die before present day assuming that the star's lifetime follows the relation \cite{CarrollEtAl2017}:
\ben
{\cal T} = 10^{10}{\rm yr}\left(\dfrac{\mstar}{\msun}\right)^{-2.5}\,.
\een
We then assign a planet to each of these stars by picking the orbital radius following a log-uniform distribution. That way, we can have a rough estimate of the expected initial distribution of planets. To introduce an artificial variance, we then compute the probability for each planet to be ejected given the DMO population, and effectively eject it if is is larger than a random number drawn between 0 and 1. Here, for simplicity, we only consider a DMO population made of compact objects like black holes, which are then characterized only by a monochromatic mass and a fraction in the Milky Way.
\begin{figure*}[!t]
\centering
\includegraphics[width=0.49\textwidth]{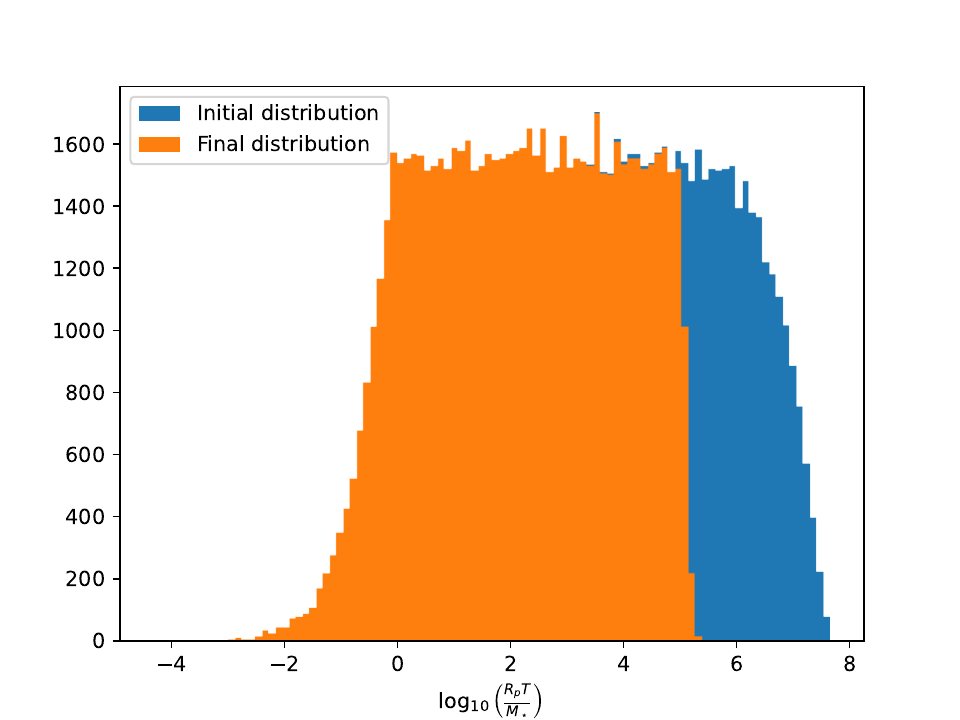}
\includegraphics[width=0.49\textwidth]{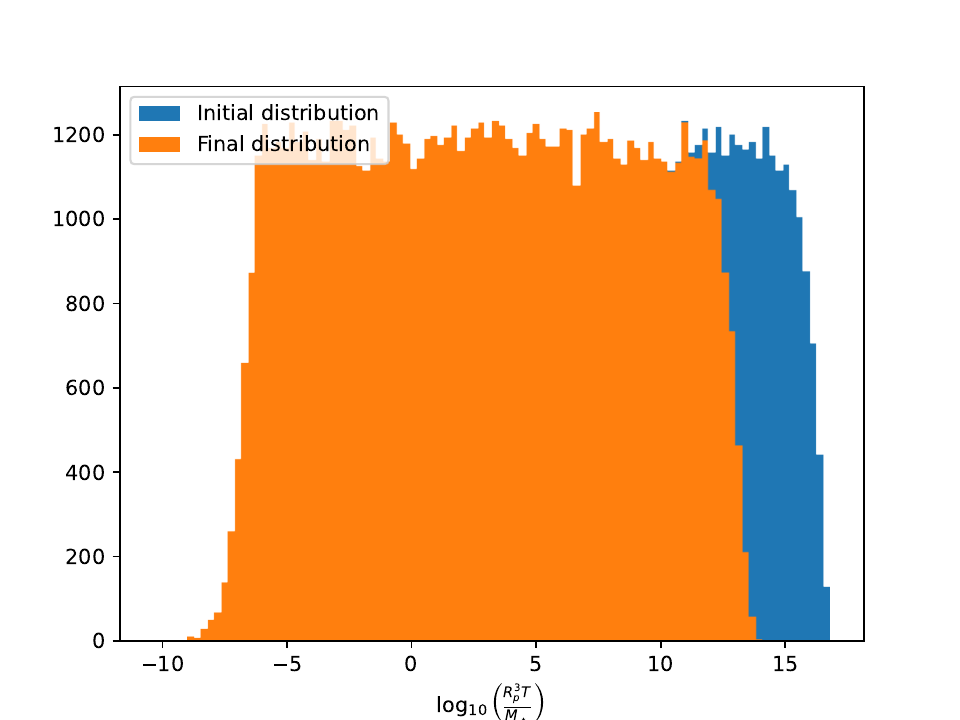}
\caption{\small {\bf Left panel}: Distribution of exoplanets as a function of $\Rplan {\cal T}/\mstar$ with and without heating from a population of compact DMOs ($\mdmo = 10^2\Msun$, $\fdmo = 1$). {\bf Right panel}: Distribution of exoplanets as a function of $\Rplan^3{\cal T}/\mstar $with and without evaporation from a population of more massive compact DMOs ($\mdmo = 10^8\Msun$, $\fdmo = 1$).}
\label{fig:histplanet_th}
\end{figure*}

The resulting distributions for $\mdmo = 10^2\Msun$ and $\mdmo = 10^8\Msun$ ($\fdmo = 1$ in both case) are shown in \citefig{fig:histplanet_th}. These distributions exhibit a clear and sharp cutoff above a given value of $\Rplan {\cal T}/\mstar$ (resp. $\Rplan^3{\cal T}/\mstar$) and no sign of a tail such as we see in \citefig{fig:histplanets}. This tends to show that our data sample has a distribution that does not depend on DMOs in the first place, and is more likely reflecting the incompleteness of current observations due to inherent sensitivity limits. This gives us confidence in the fact that our limits are not significantly biased.

\bibliographystyle{apsrev4-2}
\bibliography{exoplanetary_constraints.bib}

%\clearpage

\end{document}